\documentclass[prb,superscriptaddress,twocolumn,floatfix]{revtex4-2}
\usepackage[utf8]{inputenc}
\usepackage[T1]{fontenc}
\usepackage[english]{babel}
\usepackage{mathptmx}
\usepackage{rotating}
\usepackage{longtable}
\usepackage{listings}
\usepackage{xcolor}
\usepackage{siunitx}
\usepackage{chemformula}
\usepackage{lipsum}
\usepackage{array}
\usepackage{tikz}
\usetikzlibrary{arrows.meta}
\usepackage{wrapfig}
\definecolor{codegreen}{rgb}{0,0.6,0}
\definecolor{codegray}{rgb}{0.5,0.5,0.5}
\definecolor{codepurple}{rgb}{0.58,0,0.82}
\definecolor{backcolour}{rgb}{0.95,0.95,0.92}
\lstdefinestyle{mystyle}{
    backgroundcolor=\color{backcolour},   
    commentstyle=\color{codegreen},
    keywordstyle=\color{magenta},
    numberstyle=\tiny\color{codegray},
    stringstyle=\color{codepurple},
    basicstyle=\ttfamily\footnotesize,
    breakatwhitespace=false,         
    breaklines=true,                 
    captionpos=b,                    
    keepspaces=true,                 
    numbersep=5pt,                  
    showspaces=false,                
    showstringspaces=false,
    showtabs=false,                  
    tabsize=2
}
\PassOptionsToPackage{hyphens}{url}
\usepackage{xurl}
\usepackage{hyperref}
\hypersetup{
  colorlinks=true,
  citecolor=blue,
  filecolor=blue,
  linkcolor=blue,
  urlcolor=blue,
  breaklinks=true
}

\usepackage{times,amsmath}
\usepackage{epsfig}
\usepackage{color}
\usepackage{longtable}
\usepackage{ulem}
\usepackage{graphicx}% Include figure files
\usepackage{dcolumn}% Align table columns on decimal point
\usepackage{bm}% bold math
\usepackage{bookmark}
\usepackage{tabularx}% bold math
\usepackage{multirow}
\usepackage{booktabs}
\usepackage{tocloft}
\usepackage{etoolbox}
\apptocmd{\thebibliography}{\sloppy}{}{}
\usepackage[switch, pagewise]{lineno}
\linenumbers\relax % Commence numbering lines
\nolinenumbers

\begin{document}

\title{Benchmarking Universal Machine Learning Force Fields for Crystal Structure Prediction of High-Energy Molecular Systems}

\author{Musiha Mahfuza Mukta}
\affiliation{Department of Mechanical Engineering and Engineering Science, University of North Carolina at Charlotte, Charlotte, NC 28223, USA}

\author{Osman Goni Ridwan}
\affiliation{Department of Mechanical Engineering and Engineering Science, University of North Carolina at Charlotte, Charlotte, NC 28223, USA}

\author{Romain Perriot}
\email{rperriot@lanl.gov}
\affiliation{Theoretical Division, Los Alamos National Laboratory, Los Alamos, NM 87545, USA.}

\author{Qiang Zhu}
\email{qzhu8@charlotte.edu}
\affiliation{Department of Mechanical Engineering and Engineering Science, University of North Carolina at Charlotte, Charlotte, NC 28223, USA}
\affiliation{North Carolina Battery Complexity, Autonomous Vehicle and Electrification (BATT CAVE) Research Center, Charlotte, NC 28223, USA}

\date{\today}
\begin{abstract}
\textbf{Abstract.}
Recent developments of universal machine learning interatomic potentials (UMLIPs) offer a fast route for screening molecular crystals based on geometry relaxation and energy ranking, but their reliability across chemically diverse energetic materials remains elusive. In particular, it is unclear whether or not these UMLIPs are over-sensitive to break the desired molecular connectivity for relaxing the periodic crystals.  Herein we tested the hypothesis that classical force-field pre-relaxation can provide a more suitable starting geometry for subsequent UMLIP relaxation on a large database of high energy molecular crystals. Three models (MACE, MACE-OFF and UMA) in conjunction with the General Amber Force Field (GAFF) were applied to test this hypothesis. Among them, direct MACE-OFF and UMA showed very high relaxation success and preserved the reference geometries most closely, but they still exhibit failures for some rare cases. Using GAFF pre-relaxation can systematically reduce the number of 
failed relaxations with lower computational costs. 
Our comparative failure and robustness analyses revealed distinct trade-offs among the evaluated models. Among them, MACE-OFF achieves a better compromise between potential energy surface smoothness, structural fidelity, and stress convergence, serving as a good choice to provide a reliable foundation for automated structural optimization.
\end{abstract}

\maketitle
\makeatletter

\vspace{3mm}\noindent

\section{Introduction}
%\QZ{Here you can organize it in this way. P1. Organic crystals are important. Screening is needed. P2. Crystal structure prediction is valuable. A typical CSP needs to relax and ranking massive crystal structures. Which is traditionally done by classical FF. But the accuracy is bad. p3. Introduce the MLIPs and say how it complements the traditional CSP workflow. p4. Discuss the likely limitation. p5. introduce our work.}

Organic molecular crystals are important in many fields, including pharmaceuticals, energetic materials, organic electronics, and porous materials.~\cite{fratini2017map, friederich2019toward, saeki2019high, nematiaram2021bright, stuke2020atomic, kunkel2019finding,lin2019multifunctional,wang2018organic,morissette2004high} Their properties depend strongly on how the molecules are arranged in the crystal. Even small changes in molecular packing, hydrogen bonding, or van der Waals interactions can lead to different crystal packings with different stability, density, and physical properties (see, for instance, Refs.~\citenum{herrmann_thermal_1992, sorescu_theoretical_2010,  perriot_thermal_2022} in the case of energetic materials). Therefore, High-throughput analysis of organic crystals is valuable for understanding known materials and identifying promising new crystal forms.~\cite{Oganov-NRM-2019, Price-CSR-2014, zhu2023organic}

Crystal structure prediction (CSP) provides a practical way to explore possible molecular packings before experimental synthesis and characterization.\cite{lommerse2000test,yang2018large,QZhu-Acta-2012,Zhu-CE-2012} A typical CSP workflow generates a large number of trial structures, relaxes each structure to a nearby local minimum, and selects those lowest-energy structures for further consideration.~\cite{Oganov-Book-2011} Recently, we developed the High-Throughput Organic Crystal Structure Prediction \texttt{HTOCSP}\cite{zhu2024-htocsp} framework, which combines molecular analysis, force-field assignment, symmetric crystal generation, structure relaxation, and energy ranking in a single open-source workflow. Among these steps, geometry relaxation is one of the most important. In a traditional CSP, a large number of generated structures are usually relaxed with classical force fields because they are computationally inexpensive and can preserve the assigned molecular connectivity through explicit bond, angle, torsion, and nonbonded terms.~\cite{kim2009crystal} %HTOCSP therefore supports GAFF based relaxation through AmberTools and CHARMM. 
However, the accuracy of classical force fields can vary across different molecules or types of molecules, which may lead to errors in the relaxed geometries and final energy ranking.

Different from the classical force field, universal machine-learning interatomic potentials (UMLIP)~\cite{wines2025chips} offer a promising way to improve this stage of the CSP workflow.~\cite{Oganov-FD-2018,mukta2026structure,zhao2026integrating,li2024machine,butler2018machine,deringer2019machine} Models such as MACE, MACE-OFF, and UMA can predict energies, forces, and stresses at much lower cost than density functional theory (DFT)~\cite{becke1993density}. They can be used to perform full relaxation of both atomic positions and lattice parameters. %MACE provides broad materials coverage, MACE-OFF is designed for organic chemical environments, and UMA is trained across a wide range of molecular and materials systems. 
These models may represent better choices than classical force fields by providing a more accurate description of the final crystal geometry and energy.

However, pure UMLIP relaxation on raw structures may be problematic when the starting configuration is far from equilibrium.~\cite{chiang2025mlip} In practice, many trial structures may contain short intermolecular contacts, strained molecular geometries, poor hydrogen positions, or large initial forces. Under these conditions, UMLIPs may encounter configurations not seen during  training, resulting in potential unintended bond breaking, proton transfer, or formation of new bonds when used in relaxation.~\cite{park2026machine} Numerical convergence with respect to energy or forces alone is therefore not enough to confirm a chemically valid relaxation. 

On the other hand, the classical force fields may not excel in accurate energy prediction, but are often reliable in preserving the desired molecular connectivity, since they are based on explicit chemical bonding terms. It is therefore more prectical to avoid the exploration of unrelated potential energy surfaces caused by UMLIP's limitation by pre-relaxing the structure to configurations that maintain the targeted connectivity with classical force fields. Such pre-relaxation strategies have been recently explored for materials generation tasks with target chemical environments within the framework of AI-generative models.~\cite{ridwan2026ai, ridwan2026crystal} Similarly, it can be valuable to test if classical force fields can be used to address the limitations of pure UMLIP relaxations.

In this work, we benchmark full structural relaxation of more than 14k high-energy molecular crystal structures using three UMLIPs (MACE~\cite{Batatia2022mace}, MACE-OFF~\cite{kovacs2025mace}, and UMA~\cite{wood2026family}); both directly and in combination with the general Amber force field (GAFF)~\cite{gaff} pre-relaxation. 
This study evaluates whether current UMLIPs can robustly relax a large and chemically diverse set of high energy molecular crystals, whether classical force field preconditioning improves reliability, and the intrinsic limitations of UMLIPs applied to high-throughput organic crystal screening.

\section{Computational methodology}
Below, we outline the key components of our computational methodology. These include preparation of the molecular-crystal dataset, construction of database, implementation of the relaxation workflow within the \texttt{HTOCSP} environment, and evaluation of convergence and performance metrics.

\subsection{High energy molecular crystal dataset}
The crystal structures examined in this work were obtained from the high-explosive molecular database compiled by Davis et al.\cite{davis2024machine} The original database was constructed by mining the Cambridge Structural Database (CSD) \cite{csd} for experimentally reported single-component, solvent-free C-H-N-O compounds containing energetic functional groups, e.g. N-O bonds, azide groups, and azo groups. Structures that could not be represented using SMILES strings, polymeric compounds, free radicals, and structures measured under elevated pressure were excluded from the original study. The published database contains approximately 21,000 experimentally synthesized molecules and associated molecular and energetic-material descriptors.

In this work, all available crystal-structure files were examined individually. Records with blank, corrupted, or unreadable crystallographic information files (CIF) were removed because they could not be converted to valid periodic structures. Duplicate crystal entries were removed to prevent the same structure from being evaluated more than once. After this filtering, 14,085 unique and readable molecular crystal structures were used for all six relaxation workflows.

The original CIF files were first converted into a readable molecular crystal database. For each entry, the CSD reference code (refcode) and corresponding molecular SMILES were used to construct a \texttt{PyXtal} molecular crystal object~\cite{pyxtal} from the CIF file. If the initial structure could not be read directly, hydrogen atoms were added during a second loading attempt. The resulting structural information, molecular identity, and CSD refcode were then stored using the \texttt{PyXtal} database interface that follows the \texttt{ASE} database format.~\cite{ase}

\subsection{Interatomic Potentials and Force Fields}
\begin{enumerate}
    \item \textbf{Classical force field for pre-relaxation}. We employed GAFF for rapid molecular pre-relaxation before the UMLIP calculations. GAFF is a widely used classical force field that provides reasonable accuracy for a broad range of organic molecules. In the \texttt{HTOCSP}, parameters are assigned via \texttt{AmberTools}~\cite{amber} for GAFF, and structures are minimized with \texttt{CHARMM} \cite{charmm}. Here, molecule-specific atomic partial charges were determined using AM1-BCC method.~\cite{jakalian2000fast}

    \item \textbf{Machine-learning interatomic potentials}. We considered MACE, MACE-OFF, and UMA for full crystal relaxation, including both atomic coordinates and lattice parameters. MACE is an equivariant message-passing neural network potential that achieves near-DFT accuracy at significantly reduced computational cost, whereas MACE-OFF is a variant of MACE optimized for organic systems, and UMA is a universal potential trained across a wide range of molecular and materials systems excluding the energetic molecules. For the two stage workflows, each UMLIP was applied after GAFF pre-relaxation to obtain the final structure, energy, forces, and stress.
    The MACE calculation used the small MACE-MP model with the dispersion option enabled. MACE-OFF used the medium model. UMA used the local uma-s-1p1 checkpoint with the organic-molecular-crystal task. 
\end{enumerate}

\subsection{Relaxation workflows}
Each structure was processed through six workflows. All calculations were performed using a Python workflow implemented within the \texttt{HTOCSP} environment. In the three pure UMLIP workflows, the reference structure was relaxed with MACE, MACE-OFF, or UMA. In the two-stage workflows, the crystal was first pre-relaxed with GAFF. In the GAFF stage, only the molecular atomic positions were relaxed, and the unit cell was held fixed. The GAFF-relaxed structure was then passed to MACE, MACE-OFF, or UMA. In the final UMLIP stage, both atomic coordinates and the unit cell were optimized. Crystal symmetry was maintained with the ASE's FixSymmetry constraint~\cite{ase}. The cell and positions were optimized with the ASE UnitCellFilter and the fast inertial relaxation engine (FIRE)~\cite{bitzek2006structural}. A maximum of 5000 FIRE optimization steps were allowed and the production calculations used a target maximum force of 0.01 $\text{eV}/\text{\text{\AA}}$. The UnitCellFilter cell factor was set to 10.0. %Residual stress was not used as a stopping criterion; it was evaluated after relaxation. 
Listing~\ref{list1} provides a simplified example of using the direct MACE relaxation.

\begin{lstlisting}[language=Python, caption=Simplified example of the direct MACE relaxation workflow.,label=list1]
from ase.constraints import FixSymmetry
from ase.filters import UnitCellFilter
from ase.io import write
from ase.optimize.fire import FIRE
from mace.calculators import mace_mp
from pyxtal.db import database

refcode = "ABAGIV"
db_file = "HEM.db"
structure = database(db_file).get_pyxtal(refcode)
atoms = structure.to_ase(resort=False)
atoms.calc = mace_mp(model="small", dispersion=True, 
                     device="cpu")
atoms.set_constraint(FixSymmetry(atoms))
relax_object = UnitCellFilter(atoms, cell_factor=10.0)
optimizer = FIRE(relax_object, a=0.01, 
                 logfile=f"{refcode}_relax.log")
converged = optimizer.run(fmax=0.03, steps=5000)
write(f"{refcode}_relaxed.cif", atoms)
\end{lstlisting}

A workflow was counted as successfully relaxed when i) final UMLIP optimization reached the force criterion, and ii) the relaxed atomic coordinates could be mapped back to the reference molecular \texttt{PyXtal} representation without an invalid change in molecular connectivity. A calculation was recorded as failed when the optimizer did not finish within the allowed limit, or a valid crystal representation could not be reconstructed, or the molecular bonding pattern changed beyond the connectivity criteria via the bond length threshold values.

\section{Results and discussions}\label{bench}

\subsection{Overall relaxation statistics}
All six workflows successfully relaxed most of the structures. As shown in Table~\ref{tab:final_result}, direct MACE produced 13,749 relaxed structures and 336 failures, giving the lowest success rate of 97.61\%. GAFF pre-relaxation improved the MACE result to 13,802 relaxed structures and reduced the failures to 283. This corresponds to a success rate of 97.99\%.
MACE-OFF and UMA were more robust than MACE. Direct MACE-OFF failed for only eight structures, and GAFF+MACE-OFF reduced this number to three. Direct UMA failed for nine structures, while GAFF+UMA failed for six. The success rates for these four workflows were all greater than 99.9\%. Thus, the largest practical improvement from GAFF pre-relaxation was observed for MACE. The gain for MACE-OFF and UMA was smaller because the direct calculations already had very high completion rates.

\begin{table}[htbp]
\centering
\caption{Summary of MLIP relaxations applied to 14,085 structures.}
\label{tab:final_result}
%\begin{tabular}{p{2.6cm} p{1.4cm} p{1.4cm} p{2.0cm}}
%\begin{tabular}{p{2.6cm} p{1.4cm} p{1.4cm} p{2.0cm}}
\begin{tabular}{@{}lccc@{}}
\hline
\textbf{Method} & \textbf{Success} & \textbf{Failure} & \textbf{Success Rate} \\
\hline
MACE             & 13,749 & 336 & 97.61\% \\
GAFF+MACE        & 13,802 & 283 & 97.99\% \\
MACE-OFF         & 14,077 & 8   & 99.94\% \\
GAFF+MACE-OFF    & 14,082 & 3   & 99.98\% \\
UMA              & 14,076 & 9   & 99.94\% \\
GAFF+UMA         & 14,079 & 6   & 99.96\% \\
\hline
\end{tabular}
\end{table}

%\end{document}

After the use of GAFF-relaxed structures, the later UMLIP refinement becomes much faster and requires only about half of the pure UMLIP relaxation time. The GAFF stage was much less expensive because only the atomic positions were optimized, while the unit-cell parameters were kept fixed. Using a residual force-convergence threshold of 0.05 \text{eV}/\text{\AA}, the GAFF step required a median of only $\sim$50 s. The resulting preconditioned structures were then fully relaxed with the corresponding UMLIP, including optimization of both the atomic positions and lattice parameters. Despite the additional GAFF step, the full GAFF+UMLIP workflow remained faster than pure UMLIP relaxation.

As shown in  Table~\ref{tab:computational_cost}, GAFF+MACE required 69.5\% of the pure MACE time, GAFF+MACE-OFF required 69.7\% of the pure MACE-OFF time, and GAFF+UMA required 84.1\% of the pure UMA time. Thus, GAFF pre-relaxation reduced the median total wall time by approximately 15–30\% for all three models. These results indicate that a short, fixed cell GAFF relaxation can remove severe local strain and unfavorable atomic contacts before the more expensive full UMLIP optimization. GAFF therefore acts as a low-cost structural preconditioner, while the UMLIP remains responsible for the final relaxation of both the molecular geometry and crystal lattice.
\begin{table}[htbp]
\centering
\caption{The median computational cost (in CPU seconds) of pure and GAFF-preconditioned UMLIP relaxations. Ratio is computed as the median full GAFF+UMLIP time divided by the median pure UMLIP time.}
\label{tab:computational_cost}
\vspace{2mm}
\begin{tabular}{@{}lccc@{}}
\hline
\textbf{Method} & \textbf{Pure UMLIP} & \textbf{GAFF + UMLIP} & \textbf{Ratio}\\
\hline
MACE & 2398.9 & 52.0 + 1468.2 & 69.5\% \\
MACE-OFF & 3056.7 & 49.4 + 1942.5 & 69.7\% \\
UMA & 3143.2 & 45.0 + 2431.9 & 84.1\% \\
\hline
\end{tabular}
\end{table}

    \begin{figure*}[htbp]
    \centering
    \includegraphics[width=0.95\textwidth]{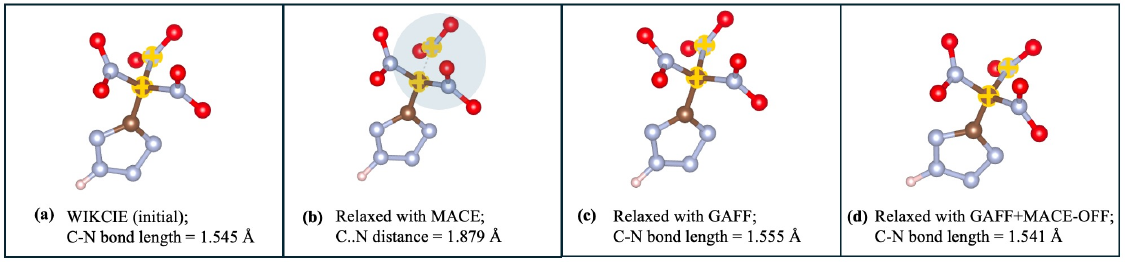}
    \vspace{-1mm}
    \caption{Representative MACE relaxation failure in trinitromethanides WIKCIE. (a) Initial structure. (b) After MACE relaxation (leading to separation of the third NO\textsubscript{2} group from molecular backbone). (c) Comparison of GAFF pre-relaxation. (d) Comparison of GAFF+MACE-OFF relaxation.}
    \label{Fig1}
    \end{figure*}

    \begin{figure*}[htbp]
    \centering
    \includegraphics[width=0.95\textwidth]{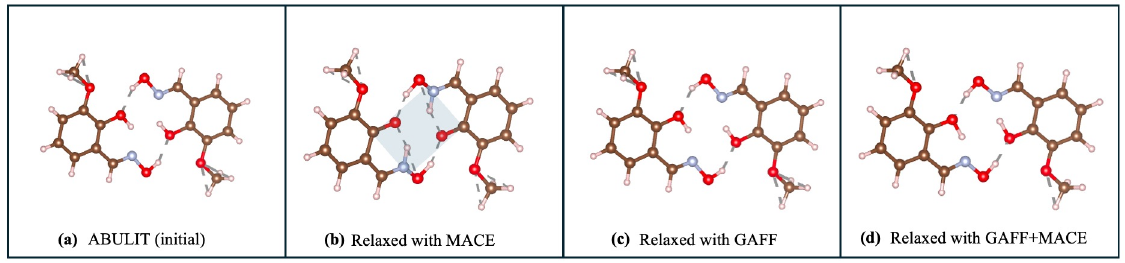}
    \vspace{-1mm}
    \caption{Representative MACE relaxation failure in ABULIT due to intramolecular proton transfer. (a) Initial structure.
    (b) After MACE relaxation (leading to intramolecular proton transfer from O-H from N-H). (c) GAFF pre-relaxation (leading to better H positions to avoid strong N-H). (d) After GAFF+MACE relaxation with the rescue.}
    \label{Fig2}
    \end{figure*}

All calculations were run on CPU nodes of the HPC cluster using Slurm. Multiple structures were relaxed in parallel within each job, with one CPU thread assigned to each relaxation worker. The reported computational costs are therefore per-structure wall-clock times.

\subsection{Failure analysis}
Across all workflows, we identified 645 method-specific failures involving 382 unique CSD refcodes. Each failure showed bond-level evidence of a molecular-connectivity change.

Among all failures, there exist 336 direct MACE and 283 GAFF+MACE cases due to inconsistent connectivity. MACE-OFF and UMA had only a small number of failures. This result is consistent with the chemical domains of the models: MACE is a general materials model, MACE-OFF is designed for neutral organic systems, and UMA contains training data from multiple chemical domains but excluding energetic crystals. Nevertheless, no model was fully protected from bond rearrangement in this chemically demanding dataset.

\begin{enumerate}

    \item \textbf{MACE}:
    MACE exhibits two types of connectivity failures. 
    First, direct MACE relaxation of trinitromethanides often results in an unphysically long bond between the central carbon and the third $\text{NO}_2$ group. Fig.~\ref{Fig1} illustrates a 
    representative example using WIKCIE. In the initial experimental structure, the $\text{C-NO}_2$ bond distance is 1.545~\AA; however, MACE relaxation leads to a significant elongation to 1.879~\AA. In contrast, GAFF is a physically motivated non-reactive force field, and it correctly yields 1.555~\AA, which is close to the experimental value. This failure likely stems from a lack of trinitromethanide configurations in the MACE training dataset, causing the model to yield nonphysical geometries. Across the entire dataset, we identified 108 trinitromethanides for which MACE consistently fails. In these cases, GAFF relaxation cannot rescue MACE, necessitating the use of alternative UMLIPs.

    In addition, another common MACE relaxation failure is the intramolecular proton transfer. Fig.~\ref{Fig2} shows that the hydrogen from the H atom from the phenolic hydroxyl group becomes attached to N site in the adjacent oxime group. This is likely because MACE may have a tendency to strongly favor the \ch{N-H} bond over \ch{O-H} bond. In this case, adding a GAFF pre-relaxation can slightly realign the initial H position to avoid connectivity. Overall, GAFF pre-relaxation rescued 82 MACE failures, including many cases where improved H alignment helped preserve \ch{X–H} connectivity.

\begin{figure*}[htbp]
    \centering
    \includegraphics[width=0.95\textwidth]{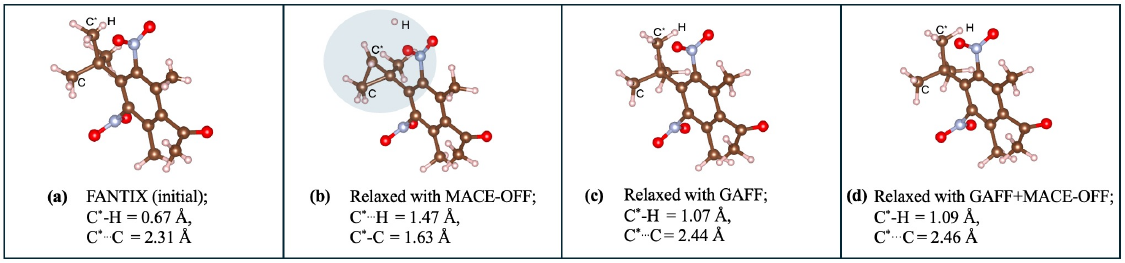}
    \vspace{-1mm}
    \caption{Example of a MACE-OFF relaxation failure rescued by GAFF pre-relaxation for FANTIX. (a) Initial structure. (b) After MACE-OFF relaxation, where the highlighted region shows a connectivity change. (c) After GAFF pre-relaxation. (d) The full GAFF+MACE-OFF workflow.}
    \label{Fig3}
\end{figure*}
    
\begin{figure*}[htbp]
    \centering
    \includegraphics[width=0.95\textwidth]{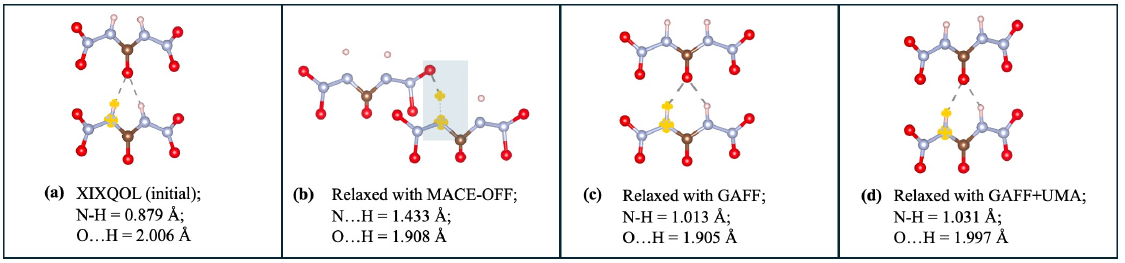}
    \vspace{-1mm}
    \caption{Example of a persistent connectivity failure in MACE-OFF relaxation for XIXQOL. (a) Initial structure. (b) MACE-OFF relaxation, where the molecule undergoes a connectivity change. (c) After GAFF pre-relaxation.  (d) An alternative UMLIP relaxation with GAFF+UMA.}
\label{Fig4}
\end{figure*}
    
    \item \textbf{MACE-OFF}: Direct MACE-OFF relaxation is much more robust than MACE, and it only yields 8 failed relaxations in the full benchmark. However, it may still have some tendency to generate artifacts for some specific configurations. %Figs.~\ref{Fig_maceoff} and ~\ref{Fig_maceoff2} illustrate two representative examples as follows. 
    
    In FANTIX (see Fig.~\ref{Fig3}), the marked methyl carbon C$^*$ initially has a rather uneven distribution of 3H+1C environment, with \ch{H-C$^*$-C} angles of 104.5$^\circ$, 112.6$^\circ$, and 117.1$^\circ$, as compared to the ideal 109.4$^\circ$.  MACE-OFF relaxation elongates one \ch{C$^*$-H} bond to 1.47 \AA\ and shortens the nearby \ch{C$^*$...C} distance from 2.31 to 1.63 \AA, producing an unintended \ch{C$^*$-C} bond and changing the coordination to 2H+2C. GAFF pre-relaxation gives more regular angles of 112.2$^\circ$, 111.0$^\circ$, and 113.3$^\circ$, preserves the \ch{C$^*$-H} bonds, and increases the non-bonded \ch{C$^*$...C} distance to 2.44 \AA. This configuration under further MACE-OFF relaxation can successfully preserve the desired connectivity. It suggests that MACE-OFF may not be capable of handling slightly distorted methyl environment and a nearby non-bonded carbon, whereas GAFF pre-relaxation provides a more suitable starting geometry that allows MACE-OFF to preserve the intended connectivity.
    
    In the case of XIXQOL in Fig. ~\ref{Fig4}, there exist hydrogen atoms lying in a sensitive intermolecular \ch{N-H...O} environment. The initial structure contains a short covalent \ch{N-H} bond of 0.879 \AA\ and a nearby contact of \ch{O...H} 2.006 \AA. During MACE-OFF relaxation, the \ch{N-H} bond elongates to 1.433 \AA, while the \ch{O...H} distance decreases only slightly to 1.908 \AA. Therefore, the hydrogen is displaced from nitrogen without forming a clear covalent \ch{O-H} bond, leaving a weakly bound local configuration. A likely reason is that MACE-OFF does not describe the balance between the covalent \ch{N-H} interaction and the neighboring intermolecular hydrogen bond accurately for this particular chemical environment. In this case, although GAFF pre-relaxation preserves the molecular connectivity through its explicit bonded terms, giving \ch{N-H}=1.013 \AA\ and maintaining the nearby \ch{O...H} contact at 1.905 \AA, a further MACE-OFF relaxation still ends up with the nonphysical \ch{O...H...N} configuration. This indicates that the problem is not caused only by an unfavorable initial geometry, but is mainly associated with the MACE-OFF description of this local \ch{N-H...O} environment. For a comparison, relaxation of the same GAFF-preconditioned structure with UMA retained the \ch{N-H} bond at 1.031 \AA\ and the intermolecular \ch{O...H} contact at 1.997 \AA. This example shows that the the UMLIP artifact caused by bond length cannot be corrected by GAFF even if it provides a better initial geometry.
    
\begin{figure*}[htbp]
    \centering
    \includegraphics[width=0.95\textwidth]{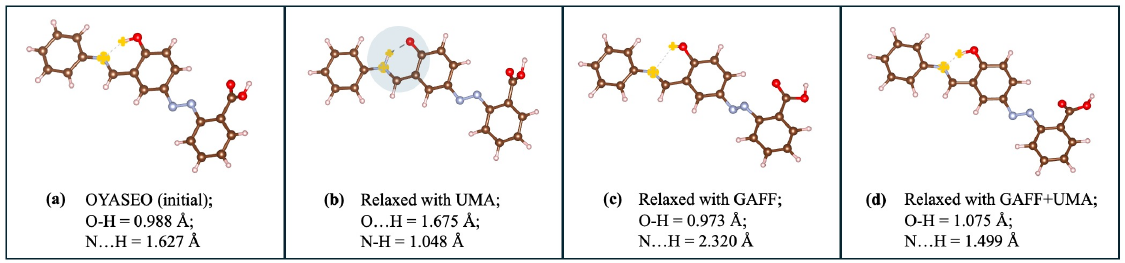}
    \vspace{-1mm}
    \caption{Example of an UMA relaxation failure rescued by GAFF pre-relaxation for OYASEO. (a) Initial structure. (b) After UMA relaxation, where the highlighted region shows O–H bond breaking and new N–H bond formation. (c) GAFF pre-relaxation. (d) GAFF+UMA workflow.}
    \label{Fig5}
\end{figure*}

\begin{figure*}[htbp]
    \centering
    \includegraphics[width=0.95\textwidth]{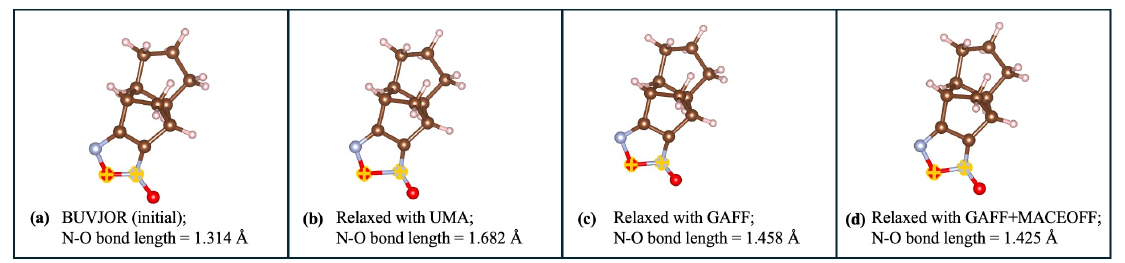}
    \vspace{-1mm}
    \caption{Example of N–O bond elongation during UMA relaxation for BUVJOR. (a) Initial structure. (b) After UMA relaxation. (c) Structure after GAFF pre-relaxation. (d) Use an alternative UMLIP relaxation GAFF+MACE-OFF to rescue the persistent UMA failure.}
    \label{Fig6}
\end{figure*}
    
    \item \textbf{UMA}: Similar to MACE-OFF, UMA is highly robust across the full benchmark, with only nine failures from direct relaxation. Note that UMA was trained on the large and diverse OMC25 dataset, potentially energetic molecules were intentionally excluded during dataset construction based on several chemical criteria, including the presence of multiple \ch{O–N}, \ch{N–N}, or \ch{O–O} bonds. Moreover, relaxation frames exhibiting changes in molecular connectivity were removed.~\cite{gharakhanyan2026open} These filtering choices improve data quality for stable molecular crystals but may reduce the representation of energetic chemical environments and bond-rearrangement regions of the potential-energy surface. This distribution difference may contribute to the few observed relaxations failures, particularly proton transfer and bond elongation in sensitive local environments, as described in the following. 
    
    In Fig.~\ref{Fig5}, the OYASEO crystal contains the short \ch{O-H...N} environment, which creates competing proton-binding sites, making the relaxation particularly sensitive to small force errors. The proton initially belongs to the hydroxyl group, with an \ch{O-H} bond length of 0.988 \AA\ and a nearby \ch{N...H} hydrogen-bond contact of 1.627 \AA. During direct UMA relaxation, the original \ch{O-H} bond is lost and the proton moves to nitrogen, forming a new \ch{N-H} bond of 1.048 \AA. GAFF pre-relaxation instead moves the proton farther from nitrogen and preserves the hydroxyl bond at 0.973 \AA. Starting from this improved local geometry, the subsequent UMA relaxation retains the \ch{O-H} connectivity, although a short \ch{N...H} hydrogen-bond contact remains. This example shows that GAFF preconditioning can correct an unfavorable proton environment and prevent UMA-driven proton transfer. 
    
    On the other hand, Fig.~\ref{Fig6} illustrates a failure that can not be rescued by GAFF pre-relaxation. In the BUVJOR crystal, a direct UMA relaxation notably increases the relevant \ch{N-O} bond increased from 1.314 to 1.682 ~\AA, which is significantly longer than a single \ch{N-O} bond of 1.3 ~\AA. Likewise, the pre-relaxation of the GAFF reduced the length of the bond to 1.458 \AA, and the subsequent relaxation of the UMA again elongated the \ch{N–O} bond to 1.682~\AA. This suggests that the UMA potential may strongly misalign this type of  \ch{N-O} single bond during its training. As a reference, MACE-OFF can correctly preserve the N–O bond at 1.425 \AA~ and produced a chemically reasonable final structure. This again exemplifies that the some UMLIP artifact related to bond length cannot be corrected by GAFF even it provides a better initial geometry.
\end{enumerate}

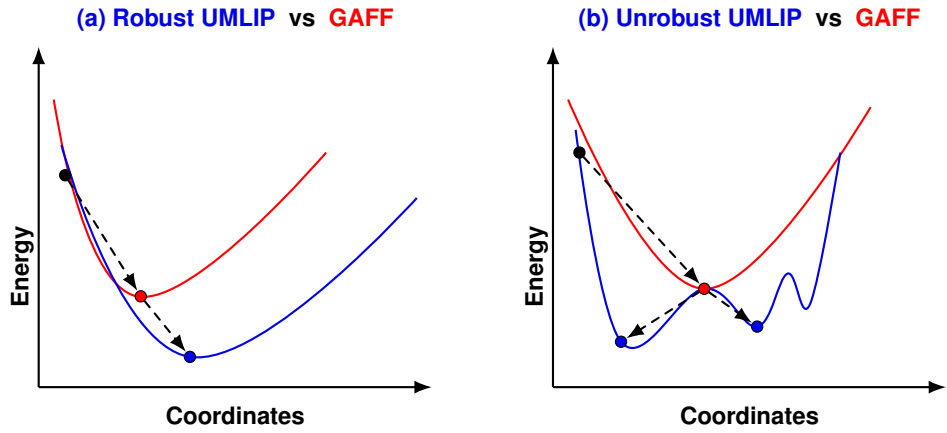
\begin{figure*}[htbp]
  \centering
  \resizebox{0.7\linewidth}{!}{%
  \begin{tikzpicture}[
      >=Latex,
      font=\sffamily\bfseries,
      % Node styles
      blkdot/.style={circle, draw=black, fill=black, inner sep=1.5pt},
      reddot/.style={circle, draw=black, fill=red, inner sep=1.5pt},
      bluedot/.style={circle, draw=black, fill=blue!90!black, inner sep=1.5pt},
      % Line styles
      gaffline/.style={draw=red, thick},
      umlipline/.style={draw=blue!90!black, thick},
      traj/.style={->, dashed, thick, line cap=round},
      axisline/.style={->, >=Latex, thick, black!60!black}
  ]

    % =========================================================
    % SUBPLOT (a): Robust UMLIP v.s GAFF
    % =========================================================
    \begin{scope}[local bounding box=panelA]
      % Title
      \node[anchor=south] at (2.6, 4.6) {\small
        \textcolor{blue!90!black}{(a) Robust UMLIP} \textcolor{black}{\text{ vs }} \textcolor{red}{GAFF}
      };

      % Coordinate Axes
      \draw[axisline] (0,0) -- (0, 4.5) node[midway, left=6pt, rotate=90, font=\sffamily\small\bfseries, black] {Energy};
      \draw[axisline] (0,0) -- (5.2, 0) node[midway, below=4pt, font=\sffamily\small\bfseries, black] {Coordinates};

      % GAFF Potential Energy Surface (Red Curve)
      \draw[gaffline] 
        plot [smooth, tension=0.75] coordinates {
          (0.2, 3.8) (1.3, 1.2) (3.8, 3.1)
        };

      % UMLIP Potential Energy Surface (Blue Curve)
      \draw[umlipline] 
        plot [smooth, tension=0.75] coordinates {
          (0.3, 3.2) (2.0, 0.4) (5.0, 2.5)
        };

      % Key Configurations / Points
      \node[blkdot] (blkA) at (0.35, 2.8) {};
      \node[reddot] (redA) at (1.35, 1.2) {};
      \node[bluedot] (blueA) at (2.0, 0.4) {};

      % Relaxation Trajectories (Dashed Arrows)
      \draw[traj] (blkA) -- (redA);
      \draw[traj] (redA) -- (blueA);
    \end{scope}

    % =========================================================
    % SUBPLOT (b): Unrobust UMLIP v.s GAFF
    % =========================================================
    \begin{scope}[xshift=6.8cm, local bounding box=panelB]
      % Title
      \node[anchor=south] at (2.6, 4.6) {\small
        \textcolor{blue!90!black}{(b) Unrobust UMLIP} \textcolor{black}{\text{ vs }} \textcolor{red}{GAFF}
      };

      % Coordinate Axes
      \draw[axisline] (0,0) -- (0, 4.5) node[midway, left=6pt, rotate=90, font=\sffamily\small\bfseries, black] {Energy};
      \draw[axisline] (0,0) -- (5.2, 0) node[midway, below=4pt, font=\sffamily\small\bfseries, black] {Coordinates};

      % GAFF Potential Energy Surface (Red Curve)
      \draw[gaffline] 
        plot [smooth, tension=0.7] coordinates {
          (0.2, 3.8) (2.0, 1.3) (4.2, 3.7)
        };

      % Unrobust UMLIP PES (Blue Multi-Well Curve with Spurious Minima)
      \draw[umlipline] 
        plot [smooth, tension=0.65] coordinates {
          (0.3, 3.4) (0.9, 0.6) (2.0, 1.3) (2.7, 0.8) (3.1, 1.5) (3.4, 1.1) (3.8, 3.1)
        };

      % Key Configurations / Points
      \node[blkdot] (blkB) at (0.35, 3.1) {};
      \node[reddot] (redB) at (2.0, 1.3) {};
      \node[bluedot] (blueB1) at (0.9, 0.6) {};
      \node[bluedot] (blueB2) at (2.7, 0.8) {};

      % Relaxation Trajectories (Dashed Arrows)
      \draw[traj] (blkB) -- (redB);
      \draw[traj] (redB) -- (blueB1);
      \draw[traj] (redB) -- (blueB2);
    \end{scope}

  \end{tikzpicture}
  }
  \caption{Comparison of relaxation pathways starting from an initial high-energy configuration (black circle) across PES described by UMLIP (blue lines) and GAFF (red lines). Dashed arrows approximates gradient descent optimization paths. (a) A robust UMLIP model provides a smooth, well-behaved PES where both the initial state and GAFF-perturbed geometry (red circle) reliably converge to the true global target minimum (blue circle). (b) An unrobust UMLIP model exhibits multiple local minima, causing GAFF-relaxed structures to get trapped in spurious local states during subsequent UMLIP optimization.}
  \label{fig7}
\end{figure*}

\subsection{UMLIP robustness}

In addition to evaluating relaxation success rates, it is important to assess how robustly each UMLIP describes the underlying potential energy surfaces (PES). Ideally, an experimentally identified configuration should correspond to a distinct PES minimum that remains smooth under small structural perturbations. During relaxation, configurations should smoothly converge to the nearest local minimum along a monotonic, energy-descending path. Although different UMLIP architectures may represent these minima with slight structural variations, a reliable model should consistently yield a well-defined unit cell characterized by low energy and minimal residual stress.

Fig.~\ref{fig7} schematically illustrates two potential relaxation outcomes. Classical force field (e.g., GAFF) optimization often serves as an intermediate pre-processing step to guide high-energy starting structures closer to the PES basin; however, it can also introduce minor perturbations to the unit cell parameters. When UMLIP relaxation is subsequently applied to these GAFF-optimized geometries, a robust UMLIP will recover the target equilibrium geometry with high fidelity, exhibiting minimal deviation in lattice constants and atomic coordinates (see Fig.~\ref{fig7}a). Conversely, an overtrained or non-robust UMLIP may generate spurious local minima and nonphysical energy barriers surrounding the true PES basin. In this scenario, the GAFF-perturbed structure can become trapped in an incorrect local state during subsequent UMLIP optimization (see Fig.~\ref{fig7}b). Therefore, evaluating UMLIP quality requires a systematic comparison of structural consistency, energy landscape smoothness, and residual stress convergence across models.

\begin{figure*}[htbp]
    \centering
    %\vspace{-3mm}
    \includegraphics[width=0.98\linewidth]{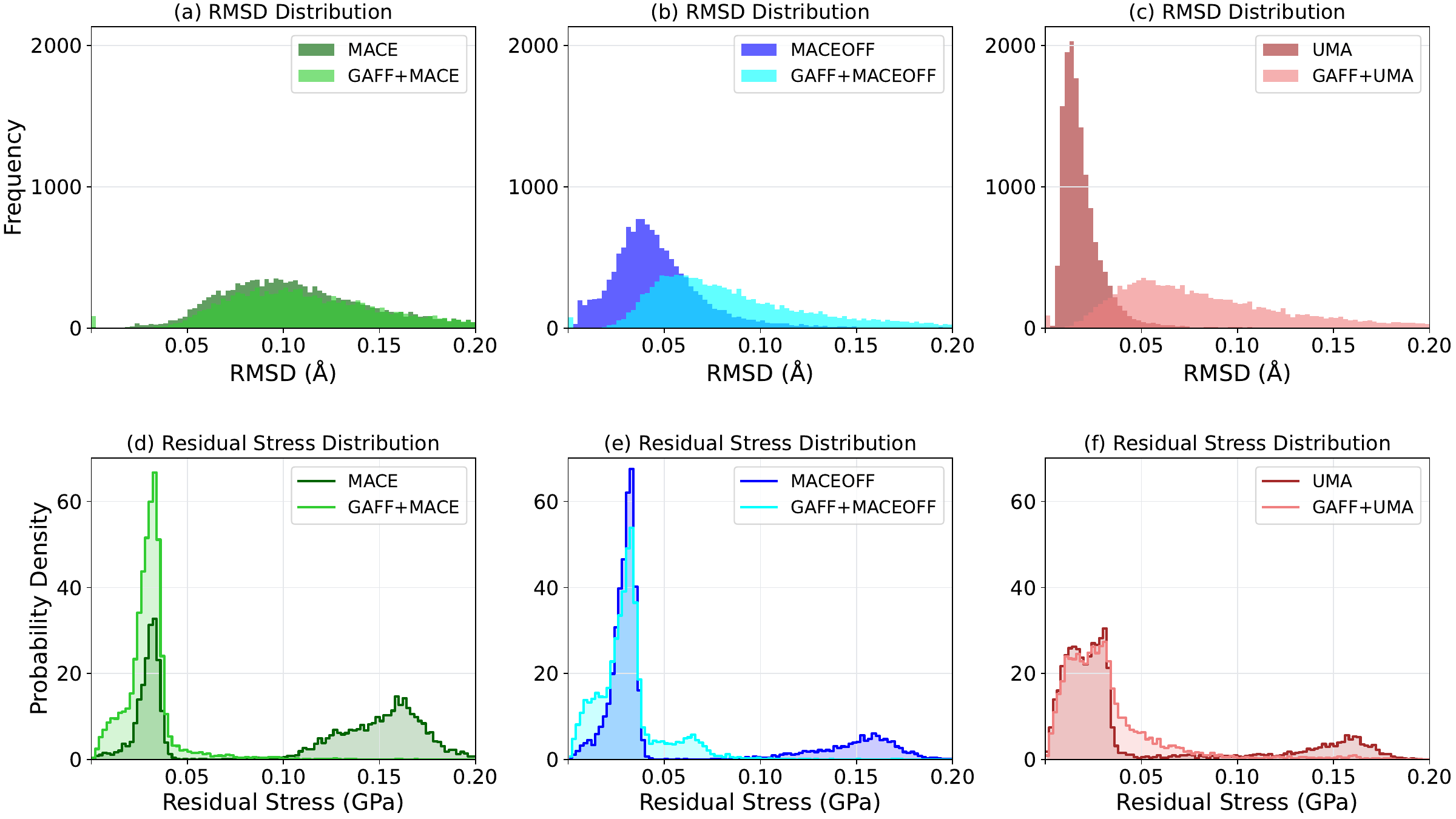}
    \caption{Comparison of structural RMSD and residual stress after pure and GAFF-preconditioned UMLIP relaxation. Panels (a–c) compare the RMSD distributions obtained with MACE, MACE-OFF, and UMA with those from their corresponding GAFF+UMLIP workflows. Panels (d–f) show the distributions of the maximum absolute residual stress-tensor component. Pure UMLIP relaxation generally produces structures closer to the reference geometry, whereas GAFF pre-relaxation substantially reduces the high-stress population.}
    \label{Fig8}
\end{figure*}

Fig.~\ref{Fig8}a compares the root-mean-square distance (RMSD) distributions of MACE-relaxed geometries relative to their initial unrelaxed structures. The strong overlap between the pure MACE and GAFF+MACE distributions indicates that MACE is robust in guiding GAFF-relaxed geometries into the true PES basin governed by a single minimum. However, the corresponding residual stress tensor distribution for MACE in Fig.~\ref{Fig8}d displays a bimodal profile, featuring a primary peak near $0.02\text{GPa}$ and a secondary peak around $0.16~\text{GPa}$. The presence of this second peak implies that a notable portion of structures fails to relax to vanishing residual stress. This incomplete stress minimization likely indicates that the MACE training objective did not place sufficient weight on stress tensors, creating ambiguity in the local stress mapping around the PES minimum.

For MACE-OFF, Fig.~\ref{Fig8}b reveals a distinct separation between the pure MACE-OFF and GAFF+MACE-OFF RMSD distributions. This divergence suggests that MACE-OFF may generate spurious local energy minima that prevent pre-relaxed geometries from reaching the same target basin. Nevertheless, the corresponding residual stress tensor plot in Fig.~\ref{Fig8}e shows a lower population in the high-stress peak ($\sim 0.16~\text{GPa}$), indicating that stress tensor training was more effectively conditioned in MACE-OFF than in standard MACE.

\begin{figure*}[t]
    \centering
    \includegraphics[width=1.0\textwidth]{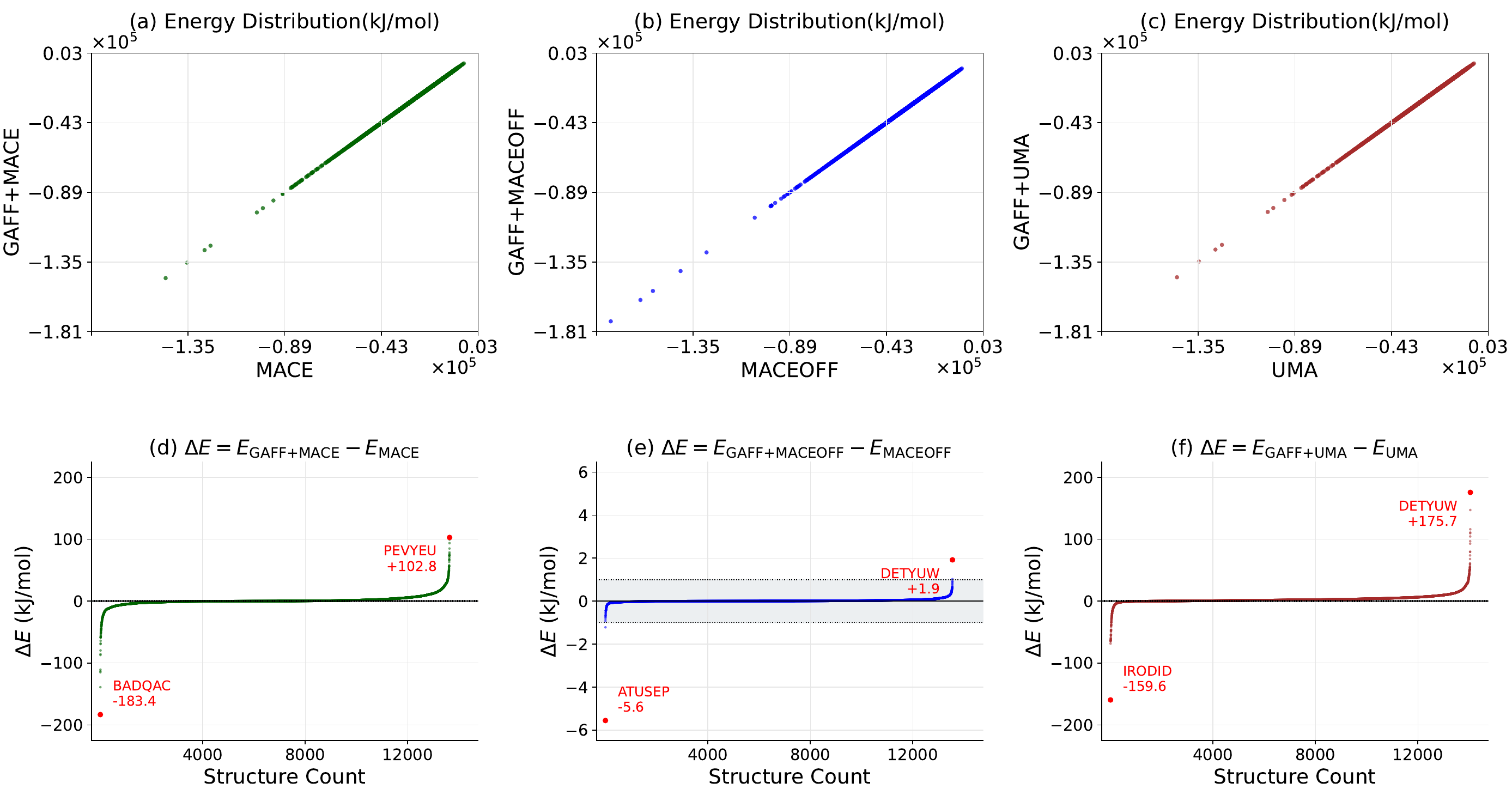}
    \vspace{-1mm}
    \caption{Comparison of final energies from pure and GAFF-preconditioned UMLIP relaxations. Panels (a–c) compare the paired final energies of the same structures evaluated with each UMLIP and its GAFF+UMLIP workflow. Panels (d–f) show the energy differences of $\Delta E = E_{\text{GAFF+UMLIP}} - E_{\text{UMLIP}}$ sorted by structure. Negative values indicate lower energies for GAFF+UMLIP.}%, whereas positive values favor UMLIP, and the shaded region marks $|\Delta E| \le 1$ kJ/mol.}
    \label{Fig9}
\end{figure*}

Compared to MACE and MACE-OFF, the UMA model exhibits the most pronounced discrepancies between pure and GAFF-assisted relaxations in RMSD distributions (see Fig.~\ref{Fig8}c), while its residual stress tensor distributions are similar to MACE-OFF (see \ref{Fig8}f). This suggests that both workflows can reach mechanically relaxed structures, but not necessarily the same structural basin. In other words, UMA appears more sensitive to the GAFF-induced perturbation of the starting geometry: after pre-relaxation, the structure may converge to a different nearby minimum instead of returning to the minimum reached by pure UMA relaxation. This behavior is qualitatively consistent with the non-robust PES scenario explained in Fig.~\ref{fig7}b, where small changes in the starting configuration can redirect the optimization toward alternative local minima.

To further confirm our analysis, we plot the comparison between pure and GAFF-preconditioned UMLIP relaxed energies in Fig. \ref{Fig9}.
At first glance, the upper panels of Fig.~\ref{Fig9} display a near-perfect one-to-one linear mapping between direct and GAFF-preconditioned relaxation energies across all three UMLIPs. This strong visual correlation is expected, given the broad energy range spanned by the over $14,000$ structure instances. 

To remove energy-scale artifacts and inspect subtle discrepancies, Figs.~\ref{Fig9}(d--f) plot the signed paired energy differences, $\Delta E = E_{\text{GAFF+UMLIP}} - E_{\text{UMLIP}}$, sorted by magnitude. For the majority of structures, $\Delta E \approx 0~\text{kJ/mol}$, indicating that both relaxation pathways converge to identical PES local minima. However, a notable divergence occurs in the tail distributions.

MACE (Fig.~\ref{Fig9}d) has substantial negative values (reaching $-183.4~\text{kJ/mol}$ for BADQAC) indicating that GAFF preconditioning assists the system in escaping shallow basins to access deeper minima. Conversely, large positive values ($+102.8~\text{kJ/mol}$ for PEVYOU) show instances where GAFF pre-relaxation diverts the trajectory away from the lower-energy minimum found via pure MACE relaxation. Similarly, UMA (Fig.~\ref{Fig9}f) displays similarly severe deviations in both directions, spanning from $-159.6~\text{kJ/mol}$ IRODID) to $+175.7~\text{kJ/mol}$ (DETYUW). In sharp contrast, MACE-OFF (see Fig.~\ref{Fig9}e) exhibits much better stability and energy consistency. Its energy differences are tightly bounded within a narrow range from $-5.6~\text{kJ/mol}$ (ATUSEP) to $+1.9~\text{kJ/mol}$ (DETYUW), with the vast majority of structures falling well within the chemically negligible $|\Delta E| \le 1.0~\text{kJ/mol}$ threshold.

In our recent work \cite{mukta2026structure}, UMA was shown to yield favorable energy agreement for near-equilibrium configurations. However, in practical organic CSP, search algorithms routinely generate a vast number of unrelaxed, high-energy trial structures that must be reliably driven toward their nearest physical PES minima. In such practical scenarios, a more robust model with a smooth and spurious-free energy landscape such as MACE-OFF represents a more reliable choice for structural optimization.

\section{Conclusions}

In this work, we benchmarked pure UMLIP and GAFF-preconditioned UMLIP relaxation workflows for over 14k high-energy molecular crystals using MACE, MACE-OFF, and UMA. All three models demonstrated high relaxation completion rates. Pre-relaxing structures with GAFF force field systematically improved optimization success by providing physically reasonable atomic alignments prior to UMLIP simulation. Furthermore, GAFF pre-conditioning significantly reduced the computational cost of subsequent UMLIP relaxations, offering a practical strategy for large-scale crystal structure workflows where UMLIP evaluation costs are prohibitive.

Our comparative failure and robustness analyzes revealed distinct trade-offs among the evaluated models. Among them, MACE is a lightweight model that efficiently captures the general PES topography, but it is more likely to yield nonphysical configurations and inaccurate energy/geometry description due to lack of training data to represent molecular geometries. However, UMA, while highly accurate for near-equilibrium energy evaluations, exhibits the largest discrepancies in both RMSD and residual stress distributions under relaxation. This behavior suggests that UMA may be susceptible to overfitting on near-equilibrium training states, generating artificial local minima and barriers when relaxing severely perturbed or high-energy structures. Despite its advantage in terms of generalization and accuracy, apply UMA for energetic crystal screening may become risky. In contrast, MACE-OFF achieves a better compromise between PES smoothness, structural fidelity, and stress convergence, serving as a superior choice to provide a reliable foundation for automated structural optimization for high throughput organic crystal screening.

\vspace{3mm}\noindent
\textbf{Author contributions}\\
Q.Z. and R.P. proposed this idea and supervised this research. M.M.M. performed the majority of materials simulations, and O.G.R constructed the online database. All coauthors designed the research, analyzed the calculations and wrote this manuscript.

\vspace{3mm}\noindent
\textbf{Conflicts of interest}\\
There are no conflicts to declare.

\vspace{3mm}\noindent
\textbf{Data availability}\\
The data and scripts used in this study, are available in \url{https://github.com/mmukta/MLIP_Becnhmark/}. All relaxed structures and associated results are stored in \url{https://mmi.charlotte.edu/hem}.

\vspace{3mm}\noindent
\textbf{\large{Acknowledgments}}\\
This project is primarily supported by the funding from the Laboratory Directed Research and Development program of Los Alamos National Laboratory under project no. 20260424ER. Los Alamos National Laboratory is operated by Triad National Security, LLC, for the National
Nuclear Security Administration of U.S. Department of Energy
(contract no. 89233218CNA000001). 
Q.Z. and M.M.M. also acknowledges the NSF (DMR-2410178) for the financial supports and 
the computing resources from ACCESS (TG-MAT230046).

\vspace{3mm}\noindent

\textbf{\large{References}}
\bibliography{ref_cleaned}

%apsrev4-2.bst 2019-01-14 (MD) hand-edited version of apsrev4-1.bst
%Control: key (0)
%Control: author (8) initials jnrlst
%Control: editor formatted (1) identically to author
%Control: production of article title (0) allowed
%Control: page (0) single
%Control: year (1) truncated
%Control: production of eprint (0) enabled
\begin{thebibliography}{47}%
\makeatletter
\providecommand \@ifxundefined [1]{%
 \@ifx{#1\undefined}
}%
\providecommand \@ifnum [1]{%
 \ifnum #1\expandafter \@firstoftwo
 \else \expandafter \@secondoftwo
 \fi
}%
\providecommand \@ifx [1]{%
 \ifx #1\expandafter \@firstoftwo
 \else \expandafter \@secondoftwo
 \fi
}%
\providecommand \natexlab [1]{#1}%
\providecommand \enquote  [1]{``#1''}%
\providecommand \bibnamefont  [1]{#1}%
\providecommand \bibfnamefont [1]{#1}%
\providecommand \citenamefont [1]{#1}%
\providecommand \href@noop [0]{\@secondoftwo}%
\providecommand \href [0]{\begingroup \@sanitize@url \@href}%
\providecommand \@href[1]{\@@startlink{#1}\@@href}%
\providecommand \@@href[1]{\endgroup#1\@@endlink}%
\providecommand \@sanitize@url [0]{\catcode `\\12\catcode `\$12\catcode
  `\&12\catcode `\#12\catcode `\^12\catcode `\_12\catcode `\%12\relax}%
\providecommand \@@startlink[1]{}%
\providecommand \@@endlink[0]{}%
\providecommand \url  [0]{\begingroup\@sanitize@url \@url }%
\providecommand \@url [1]{\endgroup\@href {#1}{\urlprefix }}%
\providecommand \urlprefix  [0]{URL }%
\providecommand \Eprint [0]{\href }%
\providecommand \doibase [0]{https://doi.org/}%
\providecommand \selectlanguage [0]{\@gobble}%
\providecommand \bibinfo  [0]{\@secondoftwo}%
\providecommand \bibfield  [0]{\@secondoftwo}%
\providecommand \translation [1]{[#1]}%
\providecommand \BibitemOpen [0]{}%
\providecommand \bibitemStop [0]{}%
\providecommand \bibitemNoStop [0]{.\EOS\space}%
\providecommand \EOS [0]{\spacefactor3000\relax}%
\providecommand \BibitemShut  [1]{\csname bibitem#1\endcsname}%
\let\auto@bib@innerbib\@empty
%</preamble>
\bibitem [{\citenamefont {Fratini}\ \emph {et~al.}(2017)\citenamefont
  {Fratini}, \citenamefont {Ciuchi}, \citenamefont {Mayou}, \citenamefont
  {De~Laissardi{\`e}re},\ and\ \citenamefont {Troisi}}]{fratini2017map}%
  \BibitemOpen
  \bibfield  {author} {\bibinfo {author} {\bibfnamefont {S.}~\bibnamefont
  {Fratini}}, \bibinfo {author} {\bibfnamefont {S.}~\bibnamefont {Ciuchi}},
  \bibinfo {author} {\bibfnamefont {D.}~\bibnamefont {Mayou}}, \bibinfo
  {author} {\bibfnamefont {G.~T.}\ \bibnamefont {De~Laissardi{\`e}re}},\ and\
  \bibinfo {author} {\bibfnamefont {A.}~\bibnamefont {Troisi}},\ }\bibfield
  {title} {\bibinfo {title} {A map of high-mobility molecular semiconductors},\
  }\href {https://doi.org/10.1038/nmat4970} {\bibfield  {journal} {\bibinfo
  {journal} {Nat. Mater.}\ }\textbf {\bibinfo {volume} {16}},\ \bibinfo {pages}
  {998} (\bibinfo {year} {2017})}\BibitemShut {NoStop}%
\bibitem [{\citenamefont {Friederich}\ \emph {et~al.}(2019)\citenamefont
  {Friederich}, \citenamefont {Fediai}, \citenamefont {Kaiser}, \citenamefont
  {Konrad}, \citenamefont {Jung},\ and\ \citenamefont
  {Wenzel}}]{friederich2019toward}%
  \BibitemOpen
  \bibfield  {author} {\bibinfo {author} {\bibfnamefont {P.}~\bibnamefont
  {Friederich}}, \bibinfo {author} {\bibfnamefont {A.}~\bibnamefont {Fediai}},
  \bibinfo {author} {\bibfnamefont {S.}~\bibnamefont {Kaiser}}, \bibinfo
  {author} {\bibfnamefont {M.}~\bibnamefont {Konrad}}, \bibinfo {author}
  {\bibfnamefont {N.}~\bibnamefont {Jung}},\ and\ \bibinfo {author}
  {\bibfnamefont {W.}~\bibnamefont {Wenzel}},\ }\bibfield  {title} {\bibinfo
  {title} {Toward design of novel materials for organic electronics},\ }\href
  {https://doi.org/10.1002/adma.201808256} {\bibfield  {journal} {\bibinfo
  {journal} {Adv. Mater.}\ }\textbf {\bibinfo {volume} {31}},\ \bibinfo {pages}
  {1808256} (\bibinfo {year} {2019})}\BibitemShut {NoStop}%
\bibitem [{\citenamefont {Saeki}\ and\ \citenamefont
  {Kranthiraja}(2019)}]{saeki2019high}%
  \BibitemOpen
  \bibfield  {author} {\bibinfo {author} {\bibfnamefont {A.}~\bibnamefont
  {Saeki}}\ and\ \bibinfo {author} {\bibfnamefont {K.}~\bibnamefont
  {Kranthiraja}},\ }\bibfield  {title} {\bibinfo {title} {A high throughput
  molecular screening for organic electronics via machine learning: present
  status and perspective},\ }\href {https://doi.org/10.7567/1347-4065/ab4f39}
  {\bibfield  {journal} {\bibinfo  {journal} {Jpn. J. Appl. Phys.}\ }\textbf
  {\bibinfo {volume} {59}},\ \bibinfo {pages} {SD0801} (\bibinfo {year}
  {2019})}\BibitemShut {NoStop}%
\bibitem [{\citenamefont {Nematiaram}\ \emph {et~al.}(2021)\citenamefont
  {Nematiaram}, \citenamefont {Padula},\ and\ \citenamefont
  {Troisi}}]{nematiaram2021bright}%
  \BibitemOpen
  \bibfield  {author} {\bibinfo {author} {\bibfnamefont {T.}~\bibnamefont
  {Nematiaram}}, \bibinfo {author} {\bibfnamefont {D.}~\bibnamefont {Padula}},\
  and\ \bibinfo {author} {\bibfnamefont {A.}~\bibnamefont {Troisi}},\
  }\bibfield  {title} {\bibinfo {title} {Bright frenkel excitons in molecular
  crystals: A survey},\ }\href {https://doi.org/10.1021/acs.chemmater.1c00645}
  {\bibfield  {journal} {\bibinfo  {journal} {Chem. Mater.}\ }\textbf {\bibinfo
  {volume} {33}},\ \bibinfo {pages} {3368} (\bibinfo {year}
  {2021})}\BibitemShut {NoStop}%
\bibitem [{\citenamefont {Stuke}\ \emph {et~al.}(2020)\citenamefont {Stuke},
  \citenamefont {Kunkel}, \citenamefont {Golze}, \citenamefont {Todorovi{\'c}},
  \citenamefont {Margraf}, \citenamefont {Reuter}, \citenamefont {Rinke},\ and\
  \citenamefont {Oberhofer}}]{stuke2020atomic}%
  \BibitemOpen
  \bibfield  {author} {\bibinfo {author} {\bibfnamefont {A.}~\bibnamefont
  {Stuke}}, \bibinfo {author} {\bibfnamefont {C.}~\bibnamefont {Kunkel}},
  \bibinfo {author} {\bibfnamefont {D.}~\bibnamefont {Golze}}, \bibinfo
  {author} {\bibfnamefont {M.}~\bibnamefont {Todorovi{\'c}}}, \bibinfo {author}
  {\bibfnamefont {J.~T.}\ \bibnamefont {Margraf}}, \bibinfo {author}
  {\bibfnamefont {K.}~\bibnamefont {Reuter}}, \bibinfo {author} {\bibfnamefont
  {P.}~\bibnamefont {Rinke}},\ and\ \bibinfo {author} {\bibfnamefont
  {H.}~\bibnamefont {Oberhofer}},\ }\bibfield  {title} {\bibinfo {title}
  {Atomic structures and orbital energies of 61,489 crystal-forming organic
  molecules},\ }\href {https://doi.org/10.1038/s41597-020-0385-y} {\bibfield
  {journal} {\bibinfo  {journal} {Sci. Data}\ }\textbf {\bibinfo {volume}
  {7}},\ \bibinfo {pages} {1} (\bibinfo {year} {2020})}\BibitemShut {NoStop}%
\bibitem [{\citenamefont {Kunkel}\ \emph {et~al.}(2019)\citenamefont {Kunkel},
  \citenamefont {Schober}, \citenamefont {Margraf}, \citenamefont {Reuter},\
  and\ \citenamefont {Oberhofer}}]{kunkel2019finding}%
  \BibitemOpen
  \bibfield  {author} {\bibinfo {author} {\bibfnamefont {C.}~\bibnamefont
  {Kunkel}}, \bibinfo {author} {\bibfnamefont {C.}~\bibnamefont {Schober}},
  \bibinfo {author} {\bibfnamefont {J.~T.}\ \bibnamefont {Margraf}}, \bibinfo
  {author} {\bibfnamefont {K.}~\bibnamefont {Reuter}},\ and\ \bibinfo {author}
  {\bibfnamefont {H.}~\bibnamefont {Oberhofer}},\ }\bibfield  {title} {\bibinfo
  {title} {Finding the right bricks for molecular legos: A data mining approach
  to organic semiconductor design},\ }\href
  {https://doi.org/10.1021/acs.chemmater.8b04436} {\bibfield  {journal}
  {\bibinfo  {journal} {Chem. Mater.}\ }\textbf {\bibinfo {volume} {31}},\
  \bibinfo {pages} {969} (\bibinfo {year} {2019})}\BibitemShut {NoStop}%
\bibitem [{\citenamefont {Lin}\ \emph {et~al.}(2019)\citenamefont {Lin},
  \citenamefont {He}, \citenamefont {Li}, \citenamefont {Wang}, \citenamefont
  {Zhou},\ and\ \citenamefont {Chen}}]{lin2019multifunctional}%
  \BibitemOpen
  \bibfield  {author} {\bibinfo {author} {\bibfnamefont {R.-B.}\ \bibnamefont
  {Lin}}, \bibinfo {author} {\bibfnamefont {Y.}~\bibnamefont {He}}, \bibinfo
  {author} {\bibfnamefont {P.}~\bibnamefont {Li}}, \bibinfo {author}
  {\bibfnamefont {H.}~\bibnamefont {Wang}}, \bibinfo {author} {\bibfnamefont
  {W.}~\bibnamefont {Zhou}},\ and\ \bibinfo {author} {\bibfnamefont
  {B.}~\bibnamefont {Chen}},\ }\bibfield  {title} {\bibinfo {title}
  {Multifunctional porous hydrogen-bonded organic framework materials},\ }\href
  {https://doi.org/10.1039/c8cs00155c} {\bibfield  {journal} {\bibinfo
  {journal} {Chem. Soc. Rev.}\ }\textbf {\bibinfo {volume} {48}},\ \bibinfo
  {pages} {1362} (\bibinfo {year} {2019})}\BibitemShut {NoStop}%
\bibitem [{\citenamefont {Wang}\ \emph {et~al.}(2018)\citenamefont {Wang},
  \citenamefont {Dong}, \citenamefont {Jiang},\ and\ \citenamefont
  {Hu}}]{wang2018organic}%
  \BibitemOpen
  \bibfield  {author} {\bibinfo {author} {\bibfnamefont {C.}~\bibnamefont
  {Wang}}, \bibinfo {author} {\bibfnamefont {H.}~\bibnamefont {Dong}}, \bibinfo
  {author} {\bibfnamefont {L.}~\bibnamefont {Jiang}},\ and\ \bibinfo {author}
  {\bibfnamefont {W.}~\bibnamefont {Hu}},\ }\bibfield  {title} {\bibinfo
  {title} {Organic semiconductor crystals},\ }\href
  {https://doi.org/10.1039/c7cs00490g} {\bibfield  {journal} {\bibinfo
  {journal} {Chem. Soc. Rev.}\ }\textbf {\bibinfo {volume} {47}},\ \bibinfo
  {pages} {422} (\bibinfo {year} {2018})}\BibitemShut {NoStop}%
\bibitem [{\citenamefont {Morissette}\ \emph {et~al.}(2004)\citenamefont
  {Morissette}, \citenamefont {Almarsson}, \citenamefont {Peterson},
  \citenamefont {Remenar}, \citenamefont {Read}, \citenamefont {Lemmo},
  \citenamefont {Ellis}, \citenamefont {Cima},\ and\ \citenamefont
  {Gardner}}]{morissette2004high}%
  \BibitemOpen
  \bibfield  {author} {\bibinfo {author} {\bibfnamefont {S.~L.}\ \bibnamefont
  {Morissette}}, \bibinfo {author} {\bibfnamefont {{\"O}.}~\bibnamefont
  {Almarsson}}, \bibinfo {author} {\bibfnamefont {M.~L.}\ \bibnamefont
  {Peterson}}, \bibinfo {author} {\bibfnamefont {J.~F.}\ \bibnamefont
  {Remenar}}, \bibinfo {author} {\bibfnamefont {M.~J.}\ \bibnamefont {Read}},
  \bibinfo {author} {\bibfnamefont {A.~V.}\ \bibnamefont {Lemmo}}, \bibinfo
  {author} {\bibfnamefont {S.}~\bibnamefont {Ellis}}, \bibinfo {author}
  {\bibfnamefont {M.~J.}\ \bibnamefont {Cima}},\ and\ \bibinfo {author}
  {\bibfnamefont {C.~R.}\ \bibnamefont {Gardner}},\ }\bibfield  {title}
  {\bibinfo {title} {High-throughput crystallization: polymorphs, salts,
  co-crystals and solvates of pharmaceutical solids},\ }\href
  {https://doi.org/10.1016/j.addr.2003.10.020} {\bibfield  {journal} {\bibinfo
  {journal} {Adv. Drug Deliv. Rev.}\ }\textbf {\bibinfo {volume} {56}},\
  \bibinfo {pages} {275} (\bibinfo {year} {2004})}\BibitemShut {NoStop}%
\bibitem [{\citenamefont {Herrmann}\ \emph {et~al.}(1992)\citenamefont
  {Herrmann}, \citenamefont {Engel},\ and\ \citenamefont
  {Eisenreich}}]{herrmann_thermal_1992}%
  \BibitemOpen
  \bibfield  {author} {\bibinfo {author} {\bibfnamefont {M.}~\bibnamefont
  {Herrmann}}, \bibinfo {author} {\bibfnamefont {W.}~\bibnamefont {Engel}},\
  and\ \bibinfo {author} {\bibfnamefont {N.}~\bibnamefont {Eisenreich}},\
  }\bibfield  {title} {{\selectlanguage {english}\bibinfo {title} {Thermal
  expansion, transitions, sensitivities and burning rates of {HMX}}},\ }\href
  {https://doi.org/10.1002/prep.19920170409} {\bibfield  {journal} {\bibinfo
  {journal} {Propellants, Explosives, Pyrotechnics}\ }\textbf {\bibinfo
  {volume} {17}},\ \bibinfo {pages} {190} (\bibinfo {year} {1992})}\BibitemShut
  {NoStop}%
\bibitem [{\citenamefont {Sorescu}\ and\ \citenamefont
  {Rice}(2010)}]{sorescu_theoretical_2010}%
  \BibitemOpen
  \bibfield  {author} {\bibinfo {author} {\bibfnamefont {D.~C.}\ \bibnamefont
  {Sorescu}}\ and\ \bibinfo {author} {\bibfnamefont {B.~M.}\ \bibnamefont
  {Rice}},\ }\bibfield  {title} {{\selectlanguage {english}\bibinfo {title}
  {Theoretical {Predictions} of {Energetic} {Molecular} {Crystals} at {Ambient}
  and {Hydrostatic} {Compression} {Conditions} {Using} {Dispersion}
  {Corrections} to {Conventional} {Density} {Functionals} ({DFT}-{D})}},\
  }\href {https://doi.org/10.1021/jp100379a} {\bibfield  {journal} {\bibinfo
  {journal} {The Journal of Physical Chemistry C}\ }\textbf {\bibinfo {volume}
  {114}},\ \bibinfo {pages} {6734} (\bibinfo {year} {2010})}\BibitemShut
  {NoStop}%
\bibitem [{\citenamefont {Perriot}\ and\ \citenamefont
  {Cawkwell}(2022)}]{perriot_thermal_2022}%
  \BibitemOpen
  \bibfield  {author} {\bibinfo {author} {\bibfnamefont {R.}~\bibnamefont
  {Perriot}}\ and\ \bibinfo {author} {\bibfnamefont {M.~J.}\ \bibnamefont
  {Cawkwell}},\ }\bibfield  {title} {{\selectlanguage {english}\bibinfo {title}
  {Thermal conductivity tensor of $\gamma$ and
  $\epsilon$-hexanitrohexaazaisowurtzitane as a function of pressure and
  temperature}},\ }\href {https://doi.org/10.1063/5.0105161} {\bibfield
  {journal} {\bibinfo  {journal} {AIP Advances}\ }\textbf {\bibinfo {volume}
  {12}},\ \bibinfo {pages} {085203} (\bibinfo {year} {2022})}\BibitemShut
  {NoStop}%
\bibitem [{\citenamefont {Oganov}\ \emph {et~al.}(2019)\citenamefont {Oganov},
  \citenamefont {Pickard}, \citenamefont {Zhu},\ and\ \citenamefont
  {Needs}}]{Oganov-NRM-2019}%
  \BibitemOpen
  \bibfield  {author} {\bibinfo {author} {\bibfnamefont {A.~R.}\ \bibnamefont
  {Oganov}}, \bibinfo {author} {\bibfnamefont {C.~J.}\ \bibnamefont {Pickard}},
  \bibinfo {author} {\bibfnamefont {Q.}~\bibnamefont {Zhu}},\ and\ \bibinfo
  {author} {\bibfnamefont {R.~J.}\ \bibnamefont {Needs}},\ }\bibfield  {title}
  {\bibinfo {title} {Structure prediction drives materials discovery},\ }\href
  {https://doi.org/10.1038/s41578-019-0101-8} {\bibfield  {journal} {\bibinfo
  {journal} {Nat. Rev. Mater.}\ }\textbf {\bibinfo {volume} {4}},\ \bibinfo
  {pages} {331} (\bibinfo {year} {2019})}\BibitemShut {NoStop}%
\bibitem [{\citenamefont {Price}(2014)}]{Price-CSR-2014}%
  \BibitemOpen
  \bibfield  {author} {\bibinfo {author} {\bibfnamefont {S.~L.}\ \bibnamefont
  {Price}},\ }\bibfield  {title} {\bibinfo {title} {Predicting crystal
  structures of organic compounds},\ }\href
  {https://doi.org/10.1039/C3CS60279F} {\bibfield  {journal} {\bibinfo
  {journal} {Chem. Soc. Rev.}\ }\textbf {\bibinfo {volume} {43}},\ \bibinfo
  {pages} {2098} (\bibinfo {year} {2014})}\BibitemShut {NoStop}%
\bibitem [{\citenamefont {Zhu}\ and\ \citenamefont
  {Hattori}(2023)}]{zhu2023organic}%
  \BibitemOpen
  \bibfield  {author} {\bibinfo {author} {\bibfnamefont {Q.}~\bibnamefont
  {Zhu}}\ and\ \bibinfo {author} {\bibfnamefont {S.}~\bibnamefont {Hattori}},\
  }\bibfield  {title} {\bibinfo {title} {Organic crystal structure prediction
  and its application to materials design},\ }\href
  {https://doi.org/10.1557/s43578-022-00698-9} {\bibfield  {journal} {\bibinfo
  {journal} {J. Mater. Res.}\ }\textbf {\bibinfo {volume} {38}},\ \bibinfo
  {pages} {19} (\bibinfo {year} {2023})}\BibitemShut {NoStop}%
\bibitem [{\citenamefont {Lommerse}\ \emph {et~al.}(2000)\citenamefont
  {Lommerse}, \citenamefont {Motherwell}, \citenamefont {Ammon}, \citenamefont
  {Dunitz}, \citenamefont {Gavezzotti}, \citenamefont {Hofmann}, \citenamefont
  {Leusen}, \citenamefont {Mooij}, \citenamefont {Price}, \citenamefont
  {Schweizer} \emph {et~al.}}]{lommerse2000test}%
  \BibitemOpen
  \bibfield  {author} {\bibinfo {author} {\bibfnamefont {J.~P.}\ \bibnamefont
  {Lommerse}}, \bibinfo {author} {\bibfnamefont {W.~S.}\ \bibnamefont
  {Motherwell}}, \bibinfo {author} {\bibfnamefont {H.~L.}\ \bibnamefont
  {Ammon}}, \bibinfo {author} {\bibfnamefont {J.~D.}\ \bibnamefont {Dunitz}},
  \bibinfo {author} {\bibfnamefont {A.}~\bibnamefont {Gavezzotti}}, \bibinfo
  {author} {\bibfnamefont {D.~W.}\ \bibnamefont {Hofmann}}, \bibinfo {author}
  {\bibfnamefont {F.~J.}\ \bibnamefont {Leusen}}, \bibinfo {author}
  {\bibfnamefont {W.~T.}\ \bibnamefont {Mooij}}, \bibinfo {author}
  {\bibfnamefont {S.~L.}\ \bibnamefont {Price}}, \bibinfo {author}
  {\bibfnamefont {B.}~\bibnamefont {Schweizer}}, \emph {et~al.},\ }\bibfield
  {title} {\bibinfo {title} {A test of crystal structure prediction of small
  organic molecules},\ }\href {https://doi.org/10.1107/S0108768100004584}
  {\bibfield  {journal} {\bibinfo  {journal} {Acta Cryst. B}\ }\textbf
  {\bibinfo {volume} {56}},\ \bibinfo {pages} {697} (\bibinfo {year}
  {2000})}\BibitemShut {NoStop}%
\bibitem [{\citenamefont {Yang}\ \emph {et~al.}(2018)\citenamefont {Yang},
  \citenamefont {De}, \citenamefont {Campbell}, \citenamefont {Li},
  \citenamefont {Ceriotti},\ and\ \citenamefont {Day}}]{yang2018large}%
  \BibitemOpen
  \bibfield  {author} {\bibinfo {author} {\bibfnamefont {J.}~\bibnamefont
  {Yang}}, \bibinfo {author} {\bibfnamefont {S.}~\bibnamefont {De}}, \bibinfo
  {author} {\bibfnamefont {J.~E.}\ \bibnamefont {Campbell}}, \bibinfo {author}
  {\bibfnamefont {S.}~\bibnamefont {Li}}, \bibinfo {author} {\bibfnamefont
  {M.}~\bibnamefont {Ceriotti}},\ and\ \bibinfo {author} {\bibfnamefont
  {G.~M.}\ \bibnamefont {Day}},\ }\bibfield  {title} {\bibinfo {title}
  {Large-scale computational screening of molecular organic semiconductors
  using crystal structure prediction},\ }\href
  {https://doi.org/10.1021/acs.chemmater.8b01621} {\bibfield  {journal}
  {\bibinfo  {journal} {Chem. Mater.}\ }\textbf {\bibinfo {volume} {30}},\
  \bibinfo {pages} {4361} (\bibinfo {year} {2018})}\BibitemShut {NoStop}%
\bibitem [{\citenamefont {Zhu}\ \emph {et~al.}(2012{\natexlab{a}})\citenamefont
  {Zhu}, \citenamefont {Oganov}, \citenamefont {Glass},\ and\ \citenamefont
  {Stokes}}]{QZhu-Acta-2012}%
  \BibitemOpen
  \bibfield  {author} {\bibinfo {author} {\bibfnamefont {Q.}~\bibnamefont
  {Zhu}}, \bibinfo {author} {\bibfnamefont {A.~R.}\ \bibnamefont {Oganov}},
  \bibinfo {author} {\bibfnamefont {C.~W.}\ \bibnamefont {Glass}},\ and\
  \bibinfo {author} {\bibfnamefont {H.~T.}\ \bibnamefont {Stokes}},\ }\bibfield
   {title} {\bibinfo {title} {Constrained evolutionary algorithm for structure
  prediction of molecular crystals: methodology and applications},\ }\href
  {https://doi.org/10.1107/S0108768112017466} {\bibfield  {journal} {\bibinfo
  {journal} {Acta Cryst. B}\ }\textbf {\bibinfo {volume} {68}},\ \bibinfo
  {pages} {215} (\bibinfo {year} {2012}{\natexlab{a}})}\BibitemShut {NoStop}%
\bibitem [{\citenamefont {Zhu}\ \emph {et~al.}(2012{\natexlab{b}})\citenamefont
  {Zhu}, \citenamefont {Oganov},\ and\ \citenamefont {Lyakhov}}]{Zhu-CE-2012}%
  \BibitemOpen
  \bibfield  {author} {\bibinfo {author} {\bibfnamefont {Q.}~\bibnamefont
  {Zhu}}, \bibinfo {author} {\bibfnamefont {A.~R.}\ \bibnamefont {Oganov}},\
  and\ \bibinfo {author} {\bibfnamefont {A.~O.}\ \bibnamefont {Lyakhov}},\
  }\bibfield  {title} {\bibinfo {title} {Evolutionary metadynamics: a novel
  method to predict crystal structures},\ }\href
  {https://doi.org/10.1039/C2CE06642D} {\bibfield  {journal} {\bibinfo
  {journal} {CrystEngComm}\ }\textbf {\bibinfo {volume} {14}},\ \bibinfo
  {pages} {3596} (\bibinfo {year} {2012}{\natexlab{b}})}\BibitemShut {NoStop}%
\bibitem [{\citenamefont {Oganov}(2011)}]{Oganov-Book-2011}%
  \BibitemOpen
  \bibfield  {author} {\bibinfo {author} {\bibfnamefont {A.~R.}\ \bibnamefont
  {Oganov}},\ }\href {https://doi.org/10.1002/9783527632831} {\emph {\bibinfo
  {title} {Modern methods of crystal structure prediction}}}\ (\bibinfo
  {publisher} {John Wiley \& Sons},\ \bibinfo {year} {2011})\BibitemShut
  {NoStop}%
\bibitem [{\citenamefont {Zhu}\ and\ \citenamefont
  {Hattori}(2025)}]{zhu2024-htocsp}%
  \BibitemOpen
  \bibfield  {author} {\bibinfo {author} {\bibfnamefont {Q.}~\bibnamefont
  {Zhu}}\ and\ \bibinfo {author} {\bibfnamefont {S.}~\bibnamefont {Hattori}},\
  }\bibfield  {title} {\bibinfo {title} {Automated high-throughput organic
  crystal structure prediction via population-based sampling},\ }\href
  {https://doi.org/10.1039/D4DD00264D} {\bibfield  {journal} {\bibinfo
  {journal} {Digital Discovery}\ }\textbf {\bibinfo {volume} {4}},\ \bibinfo
  {pages} {120} (\bibinfo {year} {2025})}\BibitemShut {NoStop}%
\bibitem [{\citenamefont {Kim}\ \emph {et~al.}(2009)\citenamefont {Kim},
  \citenamefont {Orendt}, \citenamefont {Ferraro},\ and\ \citenamefont
  {Facelli}}]{kim2009crystal}%
  \BibitemOpen
  \bibfield  {author} {\bibinfo {author} {\bibfnamefont {S.}~\bibnamefont
  {Kim}}, \bibinfo {author} {\bibfnamefont {A.~M.}\ \bibnamefont {Orendt}},
  \bibinfo {author} {\bibfnamefont {M.~B.}\ \bibnamefont {Ferraro}},\ and\
  \bibinfo {author} {\bibfnamefont {J.~C.}\ \bibnamefont {Facelli}},\
  }\bibfield  {title} {\bibinfo {title} {Crystal structure prediction of
  flexible molecules using parallel genetic algorithms with a standard force
  field},\ }\href {https://doi.org/10.1002/jcc.21189} {\bibfield  {journal}
  {\bibinfo  {journal} {J. Comput. Chem.}\ }\textbf {\bibinfo {volume} {30}},\
  \bibinfo {pages} {1973} (\bibinfo {year} {2009})}\BibitemShut {NoStop}%
\bibitem [{\citenamefont {Wines}\ and\ \citenamefont
  {Choudhary}(2025)}]{wines2025chips}%
  \BibitemOpen
  \bibfield  {author} {\bibinfo {author} {\bibfnamefont {D.}~\bibnamefont
  {Wines}}\ and\ \bibinfo {author} {\bibfnamefont {K.}~\bibnamefont
  {Choudhary}},\ }\bibfield  {title} {\bibinfo {title} {Chips-ff: Evaluating
  universal machine learning force fields for material properties},\ }\href
  {https://doi.org/10.1021/acsmaterialslett.5c00093} {\bibfield  {journal}
  {\bibinfo  {journal} {ACS Mater. Lett.}\ }\textbf {\bibinfo {volume} {7}},\
  \bibinfo {pages} {2105} (\bibinfo {year} {2025})}\BibitemShut {NoStop}%
\bibitem [{\citenamefont {Oganov}(2018)}]{Oganov-FD-2018}%
  \BibitemOpen
  \bibfield  {author} {\bibinfo {author} {\bibfnamefont {A.~R.}\ \bibnamefont
  {Oganov}},\ }\bibfield  {title} {\bibinfo {title} {Crystal structure
  prediction: reflections on present status and challenges},\ }\href
  {https://doi.org/10.1039/C8FD90033G} {\bibfield  {journal} {\bibinfo
  {journal} {Faraday Discuss.}\ }\textbf {\bibinfo {volume} {211}},\ \bibinfo
  {pages} {643} (\bibinfo {year} {2018})}\BibitemShut {NoStop}%
\bibitem [{\citenamefont {Mukta}\ \emph {et~al.}(2026)\citenamefont {Mukta},
  \citenamefont {Perriot}, \citenamefont {Hattori}, \citenamefont {Zhou},\ and\
  \citenamefont {Zhu}}]{mukta2026structure}%
  \BibitemOpen
  \bibfield  {author} {\bibinfo {author} {\bibfnamefont {M.~M.}\ \bibnamefont
  {Mukta}}, \bibinfo {author} {\bibfnamefont {R.}~\bibnamefont {Perriot}},
  \bibinfo {author} {\bibfnamefont {S.}~\bibnamefont {Hattori}}, \bibinfo
  {author} {\bibfnamefont {W.}~\bibnamefont {Zhou}},\ and\ \bibinfo {author}
  {\bibfnamefont {Q.}~\bibnamefont {Zhu}},\ }\bibfield  {title} {\bibinfo
  {title} {Structure prediction of porous organic crystals},\ }\bibfield
  {journal} {\bibinfo  {journal} {RSC Adv.}\ }\textbf {\bibinfo {volume}
  {16}},\ \href {https://doi.org/10.1039/d5ra09332e} {10.1039/d5ra09332e}
  (\bibinfo {year} {2026})\BibitemShut {NoStop}%
\bibitem [{\citenamefont {Zhao}\ \emph {et~al.}(2026)\citenamefont {Zhao},
  \citenamefont {Ma}, \citenamefont {Fan}, \citenamefont {Hu}, \citenamefont
  {Wang}, \citenamefont {Hua}, \citenamefont {Jia}, \citenamefont {Shao},
  \citenamefont {Tan}, \citenamefont {Jiang} \emph
  {et~al.}}]{zhao2026integrating}%
  \BibitemOpen
  \bibfield  {author} {\bibinfo {author} {\bibfnamefont {C.}~\bibnamefont
  {Zhao}}, \bibinfo {author} {\bibfnamefont {Z.}~\bibnamefont {Ma}}, \bibinfo
  {author} {\bibfnamefont {D.}~\bibnamefont {Fan}}, \bibinfo {author}
  {\bibfnamefont {S.}~\bibnamefont {Hu}}, \bibinfo {author} {\bibfnamefont
  {L.}~\bibnamefont {Wang}}, \bibinfo {author} {\bibfnamefont {F.}~\bibnamefont
  {Hua}}, \bibinfo {author} {\bibfnamefont {W.}~\bibnamefont {Jia}}, \bibinfo
  {author} {\bibfnamefont {E.}~\bibnamefont {Shao}}, \bibinfo {author}
  {\bibfnamefont {G.}~\bibnamefont {Tan}}, \bibinfo {author} {\bibfnamefont
  {J.}~\bibnamefont {Jiang}}, \emph {et~al.},\ }\bibfield  {title} {\bibinfo
  {title} {Integrating machine learning interatomic potentials with batched
  optimization for crystal structure prediction},\ }\href
  {https://doi.org/10.1039/d6dd00016a} {\bibfield  {journal} {\bibinfo
  {journal} {Digital Discovery}\ }\textbf {\bibinfo {volume} {5}},\ \bibinfo
  {pages} {1913} (\bibinfo {year} {2026})}\BibitemShut {NoStop}%
\bibitem [{\citenamefont {Li}\ \emph {et~al.}(2024)\citenamefont {Li},
  \citenamefont {Liang}, \citenamefont {Zhao}, \citenamefont {Wei},\ and\
  \citenamefont {Zhang}}]{li2024machine}%
  \BibitemOpen
  \bibfield  {author} {\bibinfo {author} {\bibfnamefont {C.-N.}\ \bibnamefont
  {Li}}, \bibinfo {author} {\bibfnamefont {H.-P.}\ \bibnamefont {Liang}},
  \bibinfo {author} {\bibfnamefont {B.-Q.}\ \bibnamefont {Zhao}}, \bibinfo
  {author} {\bibfnamefont {S.-H.}\ \bibnamefont {Wei}},\ and\ \bibinfo {author}
  {\bibfnamefont {X.}~\bibnamefont {Zhang}},\ }\bibfield  {title} {\bibinfo
  {title} {Machine learning assisted crystal structure prediction made
  simple},\ }\href {https://doi.org/10.20517/jmi.2024.18} {\bibfield  {journal}
  {\bibinfo  {journal} {J. Mater. Inf.}\ }\textbf {\bibinfo {volume} {4}},\
  \bibinfo {pages} {N} (\bibinfo {year} {2024})}\BibitemShut {NoStop}%
\bibitem [{\citenamefont {Butler}\ \emph {et~al.}(2018)\citenamefont {Butler},
  \citenamefont {Davies}, \citenamefont {Cartwright}, \citenamefont {Isayev},\
  and\ \citenamefont {Walsh}}]{butler2018machine}%
  \BibitemOpen
  \bibfield  {author} {\bibinfo {author} {\bibfnamefont {K.~T.}\ \bibnamefont
  {Butler}}, \bibinfo {author} {\bibfnamefont {D.~W.}\ \bibnamefont {Davies}},
  \bibinfo {author} {\bibfnamefont {H.}~\bibnamefont {Cartwright}}, \bibinfo
  {author} {\bibfnamefont {O.}~\bibnamefont {Isayev}},\ and\ \bibinfo {author}
  {\bibfnamefont {A.}~\bibnamefont {Walsh}},\ }\bibfield  {title} {\bibinfo
  {title} {Machine learning for molecular and materials science},\ }\href@noop
  {} {\bibfield  {journal} {\bibinfo  {journal} {Nature}\ }\textbf {\bibinfo
  {volume} {559}},\ \bibinfo {pages} {547} (\bibinfo {year}
  {2018})}\BibitemShut {NoStop}%
\bibitem [{\citenamefont {Deringer}\ \emph {et~al.}(2019)\citenamefont
  {Deringer}, \citenamefont {Caro},\ and\ \citenamefont
  {Cs{\'a}nyi}}]{deringer2019machine}%
  \BibitemOpen
  \bibfield  {author} {\bibinfo {author} {\bibfnamefont {V.~L.}\ \bibnamefont
  {Deringer}}, \bibinfo {author} {\bibfnamefont {M.~A.}\ \bibnamefont {Caro}},\
  and\ \bibinfo {author} {\bibfnamefont {G.}~\bibnamefont {Cs{\'a}nyi}},\
  }\bibfield  {title} {\bibinfo {title} {Machine learning interatomic
  potentials as emerging tools for materials science},\ }\href
  {https://doi.org/10.1002/adma.201902765} {\bibfield  {journal} {\bibinfo
  {journal} {Adv. Mater.}\ }\textbf {\bibinfo {volume} {31}},\ \bibinfo {pages}
  {1902765} (\bibinfo {year} {2019})}\BibitemShut {NoStop}%
\bibitem [{\citenamefont {Becke}(1993)}]{becke1993density}%
  \BibitemOpen
  \bibfield  {author} {\bibinfo {author} {\bibfnamefont {A.~D.}\ \bibnamefont
  {Becke}},\ }\bibfield  {title} {\bibinfo {title} {Density-functional
  thermochemistry. iii. the role of exact exchange},\ }\href
  {https://doi.org/10.1063/1.464913} {\bibfield  {journal} {\bibinfo  {journal}
  {J. Chem. Phys.}\ }\textbf {\bibinfo {volume} {98}},\ \bibinfo {pages} {5648}
  (\bibinfo {year} {1993})}\BibitemShut {NoStop}%
\bibitem [{\citenamefont {Chiang}\ \emph {et~al.}(2025)\citenamefont {Chiang},
  \citenamefont {Kreiman}, \citenamefont {Weaver}, \citenamefont {Amin},
  \citenamefont {Kuner}, \citenamefont {Zhang}, \citenamefont {Kaplan},
  \citenamefont {Chrzan}, \citenamefont {Blau}, \citenamefont {Krishnapriyan}
  \emph {et~al.}}]{chiang2025mlip}%
  \BibitemOpen
  \bibfield  {author} {\bibinfo {author} {\bibfnamefont {Y.}~\bibnamefont
  {Chiang}}, \bibinfo {author} {\bibfnamefont {T.}~\bibnamefont {Kreiman}},
  \bibinfo {author} {\bibfnamefont {E.}~\bibnamefont {Weaver}}, \bibinfo
  {author} {\bibfnamefont {I.}~\bibnamefont {Amin}}, \bibinfo {author}
  {\bibfnamefont {M.}~\bibnamefont {Kuner}}, \bibinfo {author} {\bibfnamefont
  {C.}~\bibnamefont {Zhang}}, \bibinfo {author} {\bibfnamefont
  {A.}~\bibnamefont {Kaplan}}, \bibinfo {author} {\bibfnamefont
  {D.}~\bibnamefont {Chrzan}}, \bibinfo {author} {\bibfnamefont {S.~M.}\
  \bibnamefont {Blau}}, \bibinfo {author} {\bibfnamefont {A.~S.}\ \bibnamefont
  {Krishnapriyan}}, \emph {et~al.},\ }\bibfield  {title} {\bibinfo {title}
  {Mlip arena: advancing fairness and transparency in machine learning
  interatomic potentials through an open and accessible benchmark platform},\
  }in\ \href@noop {} {\emph {\bibinfo {booktitle} {AI for Accelerated Materials
  Design-ICLR 2025}}}\ (\bibinfo {year} {2025})\BibitemShut {NoStop}%
\bibitem [{\citenamefont {Park}\ \emph {et~al.}(2026)\citenamefont {Park},
  \citenamefont {Pourasad}, \citenamefont {Mun}, \citenamefont {Rajan},
  \citenamefont {Park}, \citenamefont {Rohit}, \citenamefont {Zafari},
  \citenamefont {Pham}, \citenamefont {Jung}, \citenamefont {Koo} \emph
  {et~al.}}]{park2026machine}%
  \BibitemOpen
  \bibfield  {author} {\bibinfo {author} {\bibfnamefont {I.~K.}\ \bibnamefont
  {Park}}, \bibinfo {author} {\bibfnamefont {S.}~\bibnamefont {Pourasad}},
  \bibinfo {author} {\bibfnamefont {J.}~\bibnamefont {Mun}}, \bibinfo {author}
  {\bibfnamefont {A.~C.}\ \bibnamefont {Rajan}}, \bibinfo {author}
  {\bibfnamefont {T.~H.}\ \bibnamefont {Park}}, \bibinfo {author}
  {\bibfnamefont {A.}~\bibnamefont {Rohit}}, \bibinfo {author} {\bibfnamefont
  {M.}~\bibnamefont {Zafari}}, \bibinfo {author} {\bibfnamefont {T.-L.}\
  \bibnamefont {Pham}}, \bibinfo {author} {\bibfnamefont {E.}~\bibnamefont
  {Jung}}, \bibinfo {author} {\bibfnamefont {B.}~\bibnamefont {Koo}}, \emph
  {et~al.},\ }\bibfield  {title} {\bibinfo {title} {Machine learning
  interatomic potentials for energy materials: Architectures, training
  strategies, and applications},\ }\href {https://doi.org/10.1002/aenm.71046}
  {\bibfield  {journal} {\bibinfo  {journal} {Adv. Energy Mater.}\ ,\ \bibinfo
  {pages} {e71046}} (\bibinfo {year} {2026})}\BibitemShut {NoStop}%
\bibitem [{\citenamefont {Ridwan}\ \emph
  {et~al.}(2026{\natexlab{a}})\citenamefont {Ridwan}, \citenamefont
  {Piti{\'e}}, \citenamefont {Raj}, \citenamefont {Dai}, \citenamefont
  {Frapper}, \citenamefont {Xue},\ and\ \citenamefont {Zhu}}]{ridwan2026ai}%
  \BibitemOpen
  \bibfield  {author} {\bibinfo {author} {\bibfnamefont {O.~G.}\ \bibnamefont
  {Ridwan}}, \bibinfo {author} {\bibfnamefont {S.}~\bibnamefont {Piti{\'e}}},
  \bibinfo {author} {\bibfnamefont {M.~S.}\ \bibnamefont {Raj}}, \bibinfo
  {author} {\bibfnamefont {D.}~\bibnamefont {Dai}}, \bibinfo {author}
  {\bibfnamefont {G.}~\bibnamefont {Frapper}}, \bibinfo {author} {\bibfnamefont
  {H.}~\bibnamefont {Xue}},\ and\ \bibinfo {author} {\bibfnamefont
  {Q.}~\bibnamefont {Zhu}},\ }\bibfield  {title} {\bibinfo {title} {Ai-assisted
  rapid crystal structure generation towards a target local environment},\
  }\bibfield  {journal} {\bibinfo  {journal} {npj Comput. Mater.}\ }\href
  {https://doi.org/10.1038/s41524-025-01931-9} {10.1038/s41524-025-01931-9}
  (\bibinfo {year} {2026}{\natexlab{a}})\BibitemShut {NoStop}%
\bibitem [{\citenamefont {Ridwan}\ \emph
  {et~al.}(2026{\natexlab{b}})\citenamefont {Ridwan}, \citenamefont {Frapper},
  \citenamefont {Xue},\ and\ \citenamefont {Zhu}}]{ridwan2026crystal}%
  \BibitemOpen
  \bibfield  {author} {\bibinfo {author} {\bibfnamefont {O.~G.}\ \bibnamefont
  {Ridwan}}, \bibinfo {author} {\bibfnamefont {G.}~\bibnamefont {Frapper}},
  \bibinfo {author} {\bibfnamefont {H.}~\bibnamefont {Xue}},\ and\ \bibinfo
  {author} {\bibfnamefont {Q.}~\bibnamefont {Zhu}},\ }\bibfield  {title}
  {\bibinfo {title} {Crystal generation using the fully differentiable pipeline
  and latent space optimization},\ }\href
  {https://doi.org/10.1088/2632-2153/ae6751} {\bibfield  {journal} {\bibinfo
  {journal} {Mach. Learn.: Sci. Technol.}\ }\textbf {\bibinfo {volume} {7}},\
  \bibinfo {pages} {035026} (\bibinfo {year} {2026}{\natexlab{b}})}\BibitemShut
  {NoStop}%
\bibitem [{\citenamefont {Batatia}\ \emph {et~al.}(2022)\citenamefont
  {Batatia}, \citenamefont {Kovacs}, \citenamefont {Simm}, \citenamefont
  {Ortner},\ and\ \citenamefont {Cs{\'a}nyi}}]{Batatia2022mace}%
  \BibitemOpen
  \bibfield  {author} {\bibinfo {author} {\bibfnamefont {I.}~\bibnamefont
  {Batatia}}, \bibinfo {author} {\bibfnamefont {D.~P.}\ \bibnamefont {Kovacs}},
  \bibinfo {author} {\bibfnamefont {G.}~\bibnamefont {Simm}}, \bibinfo {author}
  {\bibfnamefont {C.}~\bibnamefont {Ortner}},\ and\ \bibinfo {author}
  {\bibfnamefont {G.}~\bibnamefont {Cs{\'a}nyi}},\ }\bibfield  {title}
  {\bibinfo {title} {Mace: Higher order equivariant message passing neural
  networks for fast and accurate force fields},\ }\href
  {https://doi.org/10.52202/068431-0830} {\bibfield  {journal} {\bibinfo
  {journal} {Advances in neural information processing systems}\ }\textbf
  {\bibinfo {volume} {35}},\ \bibinfo {pages} {11423} (\bibinfo {year}
  {2022})}\BibitemShut {NoStop}%
\bibitem [{\citenamefont {Kov{\'a}cs}\ \emph {et~al.}(2025)\citenamefont
  {Kov{\'a}cs}, \citenamefont {Moore}, \citenamefont {Browning}, \citenamefont
  {Batatia}, \citenamefont {Horton}, \citenamefont {Pu}, \citenamefont {Kapil},
  \citenamefont {Witt}, \citenamefont {Magdau}, \citenamefont {Cole} \emph
  {et~al.}}]{kovacs2025mace}%
  \BibitemOpen
  \bibfield  {author} {\bibinfo {author} {\bibfnamefont {D.~P.}\ \bibnamefont
  {Kov{\'a}cs}}, \bibinfo {author} {\bibfnamefont {J.~H.}\ \bibnamefont
  {Moore}}, \bibinfo {author} {\bibfnamefont {N.~J.}\ \bibnamefont {Browning}},
  \bibinfo {author} {\bibfnamefont {I.}~\bibnamefont {Batatia}}, \bibinfo
  {author} {\bibfnamefont {J.~T.}\ \bibnamefont {Horton}}, \bibinfo {author}
  {\bibfnamefont {Y.}~\bibnamefont {Pu}}, \bibinfo {author} {\bibfnamefont
  {V.}~\bibnamefont {Kapil}}, \bibinfo {author} {\bibfnamefont {W.~C.}\
  \bibnamefont {Witt}}, \bibinfo {author} {\bibfnamefont {I.-B.}\ \bibnamefont
  {Magdau}}, \bibinfo {author} {\bibfnamefont {D.~J.}\ \bibnamefont {Cole}},
  \emph {et~al.},\ }\bibfield  {title} {\bibinfo {title} {Mace-off: Short-range
  transferable machine learning force fields for organic molecules},\ }\href
  {https://doi.org/10.1021/jacs.4c07099} {\bibfield  {journal} {\bibinfo
  {journal} {J. Am. Chem. Soc.}\ }\textbf {\bibinfo {volume} {147}},\ \bibinfo
  {pages} {17598} (\bibinfo {year} {2025})}\BibitemShut {NoStop}%
\bibitem [{\citenamefont {Wood}\ \emph {et~al.}(2026)\citenamefont {Wood},
  \citenamefont {Dzamba}, \citenamefont {Fu}, \citenamefont {Gao},
  \citenamefont {Shuaibi}, \citenamefont {Barroso-Luque}, \citenamefont
  {Abdelmaqsoud}, \citenamefont {Gharakhanyan}, \citenamefont {Kitchin},
  \citenamefont {Levine} \emph {et~al.}}]{wood2026family}%
  \BibitemOpen
  \bibfield  {author} {\bibinfo {author} {\bibfnamefont {B.}~\bibnamefont
  {Wood}}, \bibinfo {author} {\bibfnamefont {M.}~\bibnamefont {Dzamba}},
  \bibinfo {author} {\bibfnamefont {X.}~\bibnamefont {Fu}}, \bibinfo {author}
  {\bibfnamefont {M.}~\bibnamefont {Gao}}, \bibinfo {author} {\bibfnamefont
  {M.}~\bibnamefont {Shuaibi}}, \bibinfo {author} {\bibfnamefont
  {L.}~\bibnamefont {Barroso-Luque}}, \bibinfo {author} {\bibfnamefont
  {K.}~\bibnamefont {Abdelmaqsoud}}, \bibinfo {author} {\bibfnamefont
  {V.}~\bibnamefont {Gharakhanyan}}, \bibinfo {author} {\bibfnamefont
  {J.}~\bibnamefont {Kitchin}}, \bibinfo {author} {\bibfnamefont
  {D.}~\bibnamefont {Levine}}, \emph {et~al.},\ }\bibfield  {title} {\bibinfo
  {title} {Uma: A family of universal models for atoms},\ }\href@noop {}
  {\bibfield  {journal} {\bibinfo  {journal} {Advances in Neural Information
  Processing Systems}\ }\textbf {\bibinfo {volume} {38}},\ \bibinfo {pages}
  {129391} (\bibinfo {year} {2026})}\BibitemShut {NoStop}%
\bibitem [{\citenamefont {Wang}\ \emph {et~al.}(2004)\citenamefont {Wang},
  \citenamefont {Wolf}, \citenamefont {Caldwell}, \citenamefont {Kollman},\
  and\ \citenamefont {Case}}]{gaff}%
  \BibitemOpen
  \bibfield  {author} {\bibinfo {author} {\bibfnamefont {J.}~\bibnamefont
  {Wang}}, \bibinfo {author} {\bibfnamefont {R.~M.}\ \bibnamefont {Wolf}},
  \bibinfo {author} {\bibfnamefont {J.~W.}\ \bibnamefont {Caldwell}}, \bibinfo
  {author} {\bibfnamefont {P.~A.}\ \bibnamefont {Kollman}},\ and\ \bibinfo
  {author} {\bibfnamefont {D.~A.}\ \bibnamefont {Case}},\ }\bibfield  {title}
  {\bibinfo {title} {Development and testing of a general amber force field},\
  }\href {https://doi.org/10.1002/jcc.20035} {\bibfield  {journal} {\bibinfo
  {journal} {J. Comput. Chem.}\ }\textbf {\bibinfo {volume} {25}},\ \bibinfo
  {pages} {1157} (\bibinfo {year} {2004})}\BibitemShut {NoStop}%
\bibitem [{\citenamefont {Davis}\ \emph {et~al.}(2024)\citenamefont {Davis},
  \citenamefont {Marrs}, \citenamefont {Cawkwell},\ and\ \citenamefont
  {Manner}}]{davis2024machine}%
  \BibitemOpen
  \bibfield  {author} {\bibinfo {author} {\bibfnamefont {J.~V.}\ \bibnamefont
  {Davis}}, \bibinfo {author} {\bibfnamefont {F.~W.}\ \bibnamefont {Marrs}},
  \bibinfo {author} {\bibfnamefont {M.~J.}\ \bibnamefont {Cawkwell}},\ and\
  \bibinfo {author} {\bibfnamefont {V.~W.}\ \bibnamefont {Manner}},\ }\bibfield
   {title} {\bibinfo {title} {Machine learning models for high explosive
  crystal density and performance},\ }\href
  {https://doi.org/10.1021/acs.chemmater.4c01978} {\bibfield  {journal}
  {\bibinfo  {journal} {Chem. Mater.}\ }\textbf {\bibinfo {volume} {36}},\
  \bibinfo {pages} {11109} (\bibinfo {year} {2024})}\BibitemShut {NoStop}%
\bibitem [{\citenamefont {Taylor}\ and\ \citenamefont {Wood}(2019)}]{csd}%
  \BibitemOpen
  \bibfield  {author} {\bibinfo {author} {\bibfnamefont {R.}~\bibnamefont
  {Taylor}}\ and\ \bibinfo {author} {\bibfnamefont {P.~A.}\ \bibnamefont
  {Wood}},\ }\bibfield  {title} {\bibinfo {title} {A million crystal
  structures: The whole is greater than the sum of its parts},\ }\href
  {https://doi.org/10.1021/acs.chemrev.9b00155} {\bibfield  {journal} {\bibinfo
   {journal} {Chem. Rev.}\ }\textbf {\bibinfo {volume} {119}},\ \bibinfo
  {pages} {9427} (\bibinfo {year} {2019})}\BibitemShut {NoStop}%
\bibitem [{\citenamefont {Fredericks}\ \emph {et~al.}(2021)\citenamefont
  {Fredericks}, \citenamefont {Parrish}, \citenamefont {Sayre},\ and\
  \citenamefont {Zhu}}]{pyxtal}%
  \BibitemOpen
  \bibfield  {author} {\bibinfo {author} {\bibfnamefont {S.}~\bibnamefont
  {Fredericks}}, \bibinfo {author} {\bibfnamefont {K.}~\bibnamefont {Parrish}},
  \bibinfo {author} {\bibfnamefont {D.}~\bibnamefont {Sayre}},\ and\ \bibinfo
  {author} {\bibfnamefont {Q.}~\bibnamefont {Zhu}},\ }\bibfield  {title}
  {\bibinfo {title} {Pyxtal: A python library for crystal structure generation
  and symmetry analysis},\ }\href {https://doi.org/10.1016/j.cpc.2020.107810}
  {\bibfield  {journal} {\bibinfo  {journal} {Comput. Phys. Commun.}\ }\textbf
  {\bibinfo {volume} {261}},\ \bibinfo {pages} {107810} (\bibinfo {year}
  {2021})}\BibitemShut {NoStop}%
\bibitem [{\citenamefont {Larsen}\ \emph {et~al.}(2017)\citenamefont {Larsen},
  \citenamefont {Mortensen}, \citenamefont {Blomqvist}, \citenamefont
  {Castelli}, \citenamefont {Christensen}, \citenamefont {Du{\l}ak},
  \citenamefont {Friis}, \citenamefont {Groves}, \citenamefont {Hammer},
  \citenamefont {Hargus} \emph {et~al.}}]{ase}%
  \BibitemOpen
  \bibfield  {author} {\bibinfo {author} {\bibfnamefont {A.~H.}\ \bibnamefont
  {Larsen}}, \bibinfo {author} {\bibfnamefont {J.~J.}\ \bibnamefont
  {Mortensen}}, \bibinfo {author} {\bibfnamefont {J.}~\bibnamefont
  {Blomqvist}}, \bibinfo {author} {\bibfnamefont {I.~E.}\ \bibnamefont
  {Castelli}}, \bibinfo {author} {\bibfnamefont {R.}~\bibnamefont
  {Christensen}}, \bibinfo {author} {\bibfnamefont {M.}~\bibnamefont
  {Du{\l}ak}}, \bibinfo {author} {\bibfnamefont {J.}~\bibnamefont {Friis}},
  \bibinfo {author} {\bibfnamefont {M.~N.}\ \bibnamefont {Groves}}, \bibinfo
  {author} {\bibfnamefont {B.}~\bibnamefont {Hammer}}, \bibinfo {author}
  {\bibfnamefont {C.}~\bibnamefont {Hargus}}, \emph {et~al.},\ }\bibfield
  {title} {\bibinfo {title} {The atomic simulation environment—a python
  library for working with atoms},\ }\href
  {https://doi.org/10.1088/1361-648X/aa680e} {\bibfield  {journal} {\bibinfo
  {journal} {J. Phys.: Condens. Matter}\ }\textbf {\bibinfo {volume} {29}},\
  \bibinfo {pages} {273002} (\bibinfo {year} {2017})}\BibitemShut {NoStop}%
\bibitem [{\citenamefont {Case}\ \emph {et~al.}(2021)\citenamefont {Case},
  \citenamefont {Aktulga}, \citenamefont {Belfon}, \citenamefont {Ben-Shalom},
  \citenamefont {Brozell}, \citenamefont {Cerutti}, \citenamefont
  {Cheatham~III}, \citenamefont {Cruzeiro}, \citenamefont {Darden},
  \citenamefont {Duke} \emph {et~al.}}]{amber}%
  \BibitemOpen
  \bibfield  {author} {\bibinfo {author} {\bibfnamefont {D.~A.}\ \bibnamefont
  {Case}}, \bibinfo {author} {\bibfnamefont {H.~M.}\ \bibnamefont {Aktulga}},
  \bibinfo {author} {\bibfnamefont {K.}~\bibnamefont {Belfon}}, \bibinfo
  {author} {\bibfnamefont {I.}~\bibnamefont {Ben-Shalom}}, \bibinfo {author}
  {\bibfnamefont {S.~R.}\ \bibnamefont {Brozell}}, \bibinfo {author}
  {\bibfnamefont {D.~S.}\ \bibnamefont {Cerutti}}, \bibinfo {author}
  {\bibfnamefont {T.~E.}\ \bibnamefont {Cheatham~III}}, \bibinfo {author}
  {\bibfnamefont {V.~W.~D.}\ \bibnamefont {Cruzeiro}}, \bibinfo {author}
  {\bibfnamefont {T.~A.}\ \bibnamefont {Darden}}, \bibinfo {author}
  {\bibfnamefont {R.~E.}\ \bibnamefont {Duke}}, \emph {et~al.},\ }\href@noop {}
  {\emph {\bibinfo {title} {Amber 2021}}}\ (\bibinfo  {publisher} {University
  of California, San Francisco},\ \bibinfo {year} {2021})\BibitemShut {NoStop}%
\bibitem [{\citenamefont {Brooks}\ \emph {et~al.}(1983)\citenamefont {Brooks},
  \citenamefont {Bruccoleri}, \citenamefont {Olafson}, \citenamefont {States},
  \citenamefont {Swaminathan},\ and\ \citenamefont {Karplus}}]{charmm}%
  \BibitemOpen
  \bibfield  {author} {\bibinfo {author} {\bibfnamefont {B.~R.}\ \bibnamefont
  {Brooks}}, \bibinfo {author} {\bibfnamefont {R.~E.}\ \bibnamefont
  {Bruccoleri}}, \bibinfo {author} {\bibfnamefont {B.~D.}\ \bibnamefont
  {Olafson}}, \bibinfo {author} {\bibfnamefont {D.~J.}\ \bibnamefont {States}},
  \bibinfo {author} {\bibfnamefont {S.}~\bibnamefont {Swaminathan}},\ and\
  \bibinfo {author} {\bibfnamefont {M.}~\bibnamefont {Karplus}},\ }\bibfield
  {title} {\bibinfo {title} {Charmm: a program for macromolecular energy,
  minimization, and dynamics calculations},\ }\href
  {https://doi.org/10.1002/jcc.540040211} {\bibfield  {journal} {\bibinfo
  {journal} {J. Comput. Chem.}\ }\textbf {\bibinfo {volume} {4}},\ \bibinfo
  {pages} {187} (\bibinfo {year} {1983})}\BibitemShut {NoStop}%
\bibitem [{\citenamefont {Jakalian}\ \emph {et~al.}(2000)\citenamefont
  {Jakalian}, \citenamefont {Bush}, \citenamefont {Jack},\ and\ \citenamefont
  {Bayly}}]{jakalian2000fast}%
  \BibitemOpen
  \bibfield  {author} {\bibinfo {author} {\bibfnamefont {A.}~\bibnamefont
  {Jakalian}}, \bibinfo {author} {\bibfnamefont {B.~L.}\ \bibnamefont {Bush}},
  \bibinfo {author} {\bibfnamefont {D.~B.}\ \bibnamefont {Jack}},\ and\
  \bibinfo {author} {\bibfnamefont {C.~I.}\ \bibnamefont {Bayly}},\ }\bibfield
  {title} {\bibinfo {title} {Fast, efficient generation of high-quality atomic
  charges. am1-bcc model: I. method},\ }\href
  {https://doi.org/10.1002/(SICI)1096-987X(20000130)21:2<132::AID-JCC5>3.0.CO;2-P}
  {\bibfield  {journal} {\bibinfo  {journal} {J. Comput. Chem.}\ }\textbf
  {\bibinfo {volume} {21}},\ \bibinfo {pages} {132} (\bibinfo {year}
  {2000})}\BibitemShut {NoStop}%
\bibitem [{\citenamefont {Bitzek}\ \emph {et~al.}(2006)\citenamefont {Bitzek},
  \citenamefont {Koskinen}, \citenamefont {G{\"a}hler}, \citenamefont
  {Moseler},\ and\ \citenamefont {Gumbsch}}]{bitzek2006structural}%
  \BibitemOpen
  \bibfield  {author} {\bibinfo {author} {\bibfnamefont {E.}~\bibnamefont
  {Bitzek}}, \bibinfo {author} {\bibfnamefont {P.}~\bibnamefont {Koskinen}},
  \bibinfo {author} {\bibfnamefont {F.}~\bibnamefont {G{\"a}hler}}, \bibinfo
  {author} {\bibfnamefont {M.}~\bibnamefont {Moseler}},\ and\ \bibinfo {author}
  {\bibfnamefont {P.}~\bibnamefont {Gumbsch}},\ }\bibfield  {title} {\bibinfo
  {title} {Structural relaxation made simple},\ }\href
  {https://doi.org/10.1103/PhysRevLett.97.170201} {\bibfield  {journal}
  {\bibinfo  {journal} {Phys. Rev. Lett.}\ }\textbf {\bibinfo {volume} {97}},\
  \bibinfo {pages} {170201} (\bibinfo {year} {2006})}\BibitemShut {NoStop}%
\bibitem [{\citenamefont {Gharakhanyan}\ \emph {et~al.}(2026)\citenamefont
  {Gharakhanyan}, \citenamefont {Barroso-Luque}, \citenamefont {Yang},
  \citenamefont {Shuaibi}, \citenamefont {Michel}, \citenamefont {Levine},
  \citenamefont {Dzamba}, \citenamefont {Fu}, \citenamefont {Gao},
  \citenamefont {Liu} \emph {et~al.}}]{gharakhanyan2026open}%
  \BibitemOpen
  \bibfield  {author} {\bibinfo {author} {\bibfnamefont {V.}~\bibnamefont
  {Gharakhanyan}}, \bibinfo {author} {\bibfnamefont {L.}~\bibnamefont
  {Barroso-Luque}}, \bibinfo {author} {\bibfnamefont {Y.}~\bibnamefont {Yang}},
  \bibinfo {author} {\bibfnamefont {M.}~\bibnamefont {Shuaibi}}, \bibinfo
  {author} {\bibfnamefont {K.}~\bibnamefont {Michel}}, \bibinfo {author}
  {\bibfnamefont {D.~S.}\ \bibnamefont {Levine}}, \bibinfo {author}
  {\bibfnamefont {M.}~\bibnamefont {Dzamba}}, \bibinfo {author} {\bibfnamefont
  {X.}~\bibnamefont {Fu}}, \bibinfo {author} {\bibfnamefont {M.}~\bibnamefont
  {Gao}}, \bibinfo {author} {\bibfnamefont {X.}~\bibnamefont {Liu}}, \emph
  {et~al.},\ }\bibfield  {title} {\bibinfo {title} {Open molecular crystals
  2025 (omc25) dataset and models},\ }\bibfield  {journal} {\bibinfo  {journal}
  {Sci. Data}\ }\href {https://doi.org/10.1038/s41597-026-06628-2}
  {10.1038/s41597-026-06628-2} (\bibinfo {year} {2026})\BibitemShut {NoStop}%
\end{thebibliography}%

\end{document}